\documentclass[12pt]{article}
\pdfoutput=1
\usepackage[DIV13]{typearea}
\usepackage{indentfirst}
\usepackage{amsmath}
\usepackage{amsfonts}
\usepackage{mathrsfs}
\usepackage{amssymb}
\usepackage{mathtools}
\usepackage{bbm}
\usepackage{epsfig}
\usepackage{graphicx}
\usepackage{subfigure}
\usepackage{slashed}
\usepackage{multicol}
\usepackage[usenames,dvipsnames]{color}
\usepackage{cite}
\RequirePackage[colorlinks=true,urlcolor=blue,anchorcolor=blue,citecolor=blue,filecolor=blue,
               linkcolor=blue,menucolor=blue,linktocpage=true,pdfproducer=medialab]{hyperref}
\usepackage{cancel}
\usepackage{feynmp}
\usepackage{lscape}
\usepackage{multirow}
\usepackage{array}
\usepackage{pdfpages}
\usepackage{fancyhdr}
\usepackage{pifont}

\usepackage[left=.9in, right=.9in]{geometry}

\newcommand{\email}[1]{\href{mailto:#1}{\tt #1}}

\numberwithin{equation}{section}
\newcommand{\LL}{\mathscr{L}}
\newcommand{\cG}{\mathscr{G}}

\def\cD{{\cal D}}
\def\cF{{\cal F}}
\def\cG{{\cal G}}
\def\cGSM{{\cal G}_{SM}}
\def\cM{{\cal M}}
\def\cN{{\cal N}}
\def\cO{{\cal O}}

\def\cP{{\cal P}}
\def\cR{{\cal R}}

\def\Tr{{\rm Tr}}

\def\cY{{\bf Y}}
\def\cW{{\cal W}}

\def\be{\begin{equation}}
\def\ee{\end{equation}}
\def\beq{\begin{equation}}
\def\eeq{\end{equation}}
\def\bc{\begin{center}}
\def\ec{\end{center}}
\def\bea{\begin{eqnarray}}
\def\eea{\end{eqnarray}}

\def\nn{\nonumber}
\def\nt{\noindent}

\renewcommand{\a}{\alpha}

\newcommand{\s}{\sigma}
\newcommand{\e}{\varepsilon}

\newcommand{\derp}{\partial}

\newcommand{\hc}{\mathrm{h.c.}}

\newcommand{\LT}{\mathbf{L}}
\newcommand{\RT}{\mathbf{R}}
\newcommand{\UY}{\mathbf{U_Y}}
\newcommand{\UH}{\mathbf{U}}
\newcommand{\UHL}{\mathbf{U}_L}
\newcommand{\UHR}{\mathbf{U}_R}
\newcommand{\UHLR}{\mathbf{U}_{L(R)}}

\newcommand{\TL}{\mathbf{T}}
\newcommand{\VL}{\mathbf{V}}
\newcommand{\DL}{D}

\newcommand{\DLL}{\mathcal{D}}

\newcommand{\TLL}{\mathbf{T}_L}
\newcommand{\TLR}{\mathbf{T}_R}

\newcommand{\VLmuu}{\mathbf{V}^\mu}
\newcommand{\VLmud}{\mathbf{V}_{\mu}}

\newcommand{\VLLmuu}{\mathbf{V}^\mu_L}
\newcommand{\VLLmud}{\mathbf{V}_{\mu,\,L}}
\newcommand{\VLLnuu}{\mathbf{V}^\nu_L}

\newcommand{\VLRmuu}{\mathbf{V}^\mu_R}
\newcommand{\VLRmud}{\mathbf{V}_{\mu,\,R}}
\newcommand{\VLRnuu}{\mathbf{V}^\nu_R}

\newcommand{\VLLmuut}{\widetilde{\mathbf{V}}^\mu_L}
\newcommand{\VLLmudt}{\widetilde{\mathbf{V}}_{\mu,\,L}}
\newcommand{\VLLnuut}{\widetilde{\mathbf{V}}^\nu_L}
\newcommand{\VLLnudt}{\widetilde{\mathbf{V}}_{\nu,\,L}}
\newcommand{\VLRmuut}{\widetilde{\mathbf{V}}^\mu_R}
\newcommand{\VLRmudt}{\widetilde{\mathbf{V}}_{\mu,\,R}}
\newcommand{\VLRnuut}{\widetilde{\mathbf{V}}^\nu_R}
\newcommand{\VLRnudt}{\widetilde{\mathbf{V}}_{\nu,\,R}}

\newcommand{\WWLut}{\widetilde{W}^{\mu\nu}_L}
\newcommand{\WWRut}{\widetilde{W}^{\mu\nu}_R}

\newcommand{\WWRdt}{\widetilde{W}_{\mu\nu,\,R}}

\newcommand{\WWWLut}{\widetilde{W}^{\rho\sigma}_L}
\newcommand{\WWWRut}{\widetilde{W}^{\rho\sigma}_R}

\newcommand{\TLchi}{\mathbf{T}_\chi}
\newcommand{\VLchimuu}{\mathbf{V}^\mu_\chi}
\newcommand{\VLchimud}{\mathbf{V}_{\mu,\,\chi}}
\newcommand{\VLchinuu}{\mathbf{V}^\nu_\chi}
\newcommand{\VLchinud}{\mathbf{V}_{\nu,\,\chi}}

\newcommand{\WWchiu}{W^{\mu\nu}_\chi}

\newcommand{\WWchiut}{\widetilde{W}^{\mu\nu}_\chi}
\newcommand{\WWWchiu}{W^{\rho\sigma}_\chi}

\newcommand{\TLLR}{\mathbf{T}_{L(R)}}
\newcommand{\VLLRmuu}{\mathbf{V}^\mu_{L(R)}}
\newcommand{\VLLRnuu}{\mathbf{V}^\nu_{L(R)}}
\newcommand{\TLLt}{\widetilde{\mathbf{T}}_{L}}
\newcommand{\TLRt}{\widetilde{\mathbf{T}}_{R}}

\newcommand{\VLLRmuut}{\widetilde{\mathbf{V}}^\mu_{L(R)}}
\newcommand{\VLRLmuut}{\widetilde{\mathbf{V}}^\mu_{R(L)}}
\newcommand{\VLRLmudt}{\widetilde{\mathbf{V}}_{\mu,R(L)}}
\newcommand{\VLRLnuut}{\widetilde{\mathbf{V}}^\nu_{R(L)}}

\newcommand{\WWLRu}{W^{\mu\nu}_{L(R)}}

\newcommand{\WWRLut}{\widetilde{W}^{\mu\nu}_{R(L)}}

\newcommand{\fL}{f_L}
\newcommand{\fR}{f_R}
\newcommand{\fLR}{f_{L(R)}}
\newcommand{\fRL}{f_{R(L)}}
\newcommand{\fchi}{f_\chi}
\newcommand{\gL}{g_L}
\newcommand{\gR}{g_R}
\newcommand{\gLR}{g_{L(R)}}

\newcommand{\gchi}{g_\chi}
\newcommand{\GF}{G_F}
\newcommand{\aem}{\alpha_\text{em}}

\newcommand{\tr}{\Tr}

\renewcommand{\to}{\rightarrow}

\newcommand{\BBu}{B^{\mu\nu}}
\newcommand{\BBd}{B_{\mu\nu}}
\newcommand{\WWu}{W^{\mu\nu}}

\newcommand{\bmat}{\begin{pmatrix}}
\newcommand{\emat}{\end{pmatrix}}

\newcommand{\brown}[1]{\color[rgb]{0.8,0.1,0.} #1 \color{black}}

\begin{document}
\begin{titlepage}
\vspace*{-1cm}
\phantom{hep-ph/***} 
{\flushleft
\hfill{}\\
\hfill{}}
\vskip 1cm
\begin{center}
\mathversion{bold}
{\LARGE\bf Left-right non-linear dynamical Higgs}\\
\mathversion{normal}
\vskip .3cm
\end{center}
\vskip 0.5  cm
\begin{center}
{\large Jing Shu}~$^{a)}$,
{\large Juan Yepes}~$^{a)}$
\\
\vskip .7cm
{\footnotesize
$^{a)}$~
State Key Laboratory of Theoretical Physics and Kavli Institute for Theoretical Physics China,\\
Institute of Theoretical Physics, Chinese Academy of Sciences, Beijing 100190, P. R. China\\
\vskip .1cm
\vskip .3cm
\begin{minipage}[l]{.9\textwidth}
\begin{center} 
\textit{E-mail:} 
\email{juyepes@itp.ac.cn},
\email{jshu@itp.ac.cn}
\end{center}
\end{minipage}
}
\end{center}
\vskip 0.5cm
\begin{abstract}
All the possible CP-conserving non-linear operators up to the $p^4$-order in the Lagrangian expansion are analysed here for the left-right symmetric model in the non-linear electroweak chiral context coupled to a light dynamical Higgs. The low energy effects will be triggered by  an emerging new physics field content in the nature, more specifically, from spin-1 resonances sourced by the straightforward extension of the SM local gauge symmetry to the larger local group $SU(2)_L\otimes SU(2)_R\otimes U(1)_{B-L}$. Low energy phenomenology will be altered by integrating out the resonances from the physical spectrum, being manifested through induced corrections onto the left handed operators.  Such modifications are weighted by powers of the scales ratio implied by the symmetries of the model  and will determine the size of the effective operator basis to be used. The recently observed diboson excess around the invariant mass 1.8--2 TeV entails a scale suppression that suggests to encode the low energy effects via a much smaller set of effective operators.

\end{abstract}
\end{titlepage}
\setcounter{footnote}{0}

\tableofcontents

%
%

\newpage

%
%
\section{Introduction}

\nt A new scalar resonance~\cite{:2012gk,:2012gu} has been discovered in our nature at the LHC and experimentally confirmed as a particle resembling the Higgs boson~\cite{Englert:1964et,Higgs:1964ia,Higgs:1964pj}, establishing thus the Standard Model (SM) as a successful and consistent framework of electroweak symmetry breaking (EWSB). The role of the Higgs particle in the EWSB mechanism signals different BSM scenarios. In one class of models, the Higgs is just an elementary scalar doublet linearly transforming under the SM gauge group $SU(2)_L \otimes U(1)_Y$. Alternatively, the Higgs particle could emerge from a given strong dynamics at the TeV or slightly higher scale, and arising either as an EW doublet or as a member of other representations.  Both cases call for new physics (NP) at the TeV scale and tend to propose the existence of lighter exotic resonances which have failed to show up in data so far.

The alternative case assumes a non-perturbative Higgs dynamics associated to a strong interacting sector at $\Lambda_s$-scale, explicitly implementing a non--linear symmetry in the scalar sector, and sharing thus a reminiscence of the long ago proposed ``Technicolor" formalism \cite{Susskind:1978ms,Dimopoulos:1979es,Dimopoulos:1981xc}. No Higgs particle was present in the physical spectrum of such scenarios, only three would-be-Goldstone bosons (GB) were playing a role and with an associated scale $f$ identified with the electroweak scale $f=v\equiv246$ GeV (respecting $f \ge \Lambda_s/4 \pi$~\cite{Manohar:1983md}), responsible a posteriori for the weak gauge boson masses. There has been a revival in this direction relying on the fact that the Higgs particle $h$ may be light as being itself a GB	 resulting from the spontaneous breaking of a strong dynamics with symmetry group  $G$ at the scale $\Lambda_s$~\cite{Kaplan:1983fs,Kaplan:1983sm,Banks:1984gj,
Georgi:1984ef,Georgi:1984af,Dugan:1984hq}.  A subsequent source of explicit breaking of $G$ would allow the Higgs boson to pick a small mass, much as the pion gets a mass in QCD, and developing a potential with a non-trivial minimum $\langle h\rangle$. Via this explicit breaking the EW gauge symmetry is broken and the electroweak scale $v$ generated, distinct from $f$. Three scales, in addition to the strong interacting scale $\Lambda_s$, enter thus into the scenario now: $f$, $v$ and $\langle h\rangle$, although a model-dependent constraint will link them. The non--linearity strength is quantified by the ratio $v^2/f^2$, such that $f\sim v$ characterizes non--linear constructions, whilst $f\gg v$ labels  regimes approaching the linear one. 

In this work an EW strongly interacting sector coupled to the light Higgs particle will be assumed. Furthermore, motivated by the high energy regimes reachable at the LHC and future colliders, this work faces the hypothetical situation of non--zero signals arising out from some emerging new physics field content in the nature, more specifically, from spin--1 resonances driven by extending the SM local gauge symmetry $\cG_{SM}=SU(2)_L\otimes U(1)_Y$ to the larger local group $\cG=SU(2)_L\otimes SU(2)_R\otimes U(1)_{B-L}$ (see~\cite{LRSM1,LRSM2} for left-right symmetric models literature). The underlying framework employs a non--linear $\sigma$--model for the strong dynamics giving rise to the GB, i.e. the $W^\pm_L$ and $Z_L$ longitudinal components that leads to introduce the Goldstone scale $\fL$, together with the corresponding GB from the extended local group, the additional $W^\pm_R$ and $Z_R$ longitudinal degrees of freedom and the associated Goldstone scale $\fR$. Their transformation are customarily parametrized via the dimensionless unitary matrix $\UH(x)$, more specifically through $\UHL(x)$ and $\UHR(x)$ for the symmetry group $SU(2)_L\otimes SU(2)_R$, and defined as
\beq
\UHLR\,(x)=e^{i\,\tau_a\,\pi^a_{L(R)}(x)/\fLR}\, , 
\label{Goldstone-matrices}
\eeq

\nt with $\pi^a_{L(R)}(x)$ the corresponding Goldstone bosons fields suppressed by their associated non--linear sigma model scale $\fLR$. Finally, this non--linear effective set--up is coupled a posteriori to a Higgs scalar singlet $h$ in a general way through powers of $h/\fL$~\cite{Georgi:1984af}, via the generic light Higgs polynomial functions $\cF(h)$~\cite{Alonso:2012px} expandable as
\beq
\cF_i(h)\equiv1+2\,{a}_i\,\frac{h}{\fL}+{b}_i\,\frac{h^2}{\fL^2}+\cO\Big(\frac{h^3}{\fL^3}\Big)\,.
\label{F}
\eeq

\nt The whole tower of linearly independent left, right and the interplaying left--right handed operators (LH, RH and LRH respectively) for the CP--conserving bosonic sector has been established\footnote{This work completes the basis given in~\cite{Zhang:2007xy,Wang:2008nk} for the left--right symmetric EW chiral models, it generalizes the work done in~\cite{Appelquist:1980vg,Longhitano:1980iz,Longhitano:1980tm,
Feruglio:1992wf,Appelquist:1993ka} for the heavy Higgs chiral scenario, and it extends as well the dynamical Higgs scenario~\cite{Alonso:2012px,Brivio:2013pma} to the case of a larger local gauge symmetry $\cG$ in the context of non--linear EW interactions coupled to a light Higgs particle. The CP--violating counterpart has been analysed in~\cite{Yepes:2015qwa}.} through~\cite{Yepes:2015zoa} and it is summarized again for the purposes of our left-right model analysed here. Such scenario may be considered as a generic UV completion of the low energy non--linear treatment of Refs.~\cite{Appelquist:1980vg,Longhitano:1980iz,Longhitano:1980tm,
Feruglio:1992wf,Appelquist:1993ka}  and~\cite{Alonso:2012px,Brivio:2013pma}. Its physical impact has been studied by integrating out the right handed gauge sector from the physical spectrum, leading the RH and the mixing LRH operators to collapse directly onto the LH sector, and inducing therefore corrections weighted by powers of the parameter $\bar{\epsilon}\equiv \epsilon\,c_{C,LR}$, with the scale ratio $\epsilon\equiv \fL/\fR$ and the coefficient $c_{C,LR}$ encoding the strength of the mixing among the LH and RH gauge masses. This feature leads to modify, consequently, the electroweak precision data (EWPD) parameters, the triple gauge couplings (TGC), $hVV$--couplings and the anomalous quartic gauge couplings (QGC). Corresponding allowed ranges for the involved coefficients will be also reported. 

The recently observed diboson excess at the ATLAS and CMS Collaborations around the invariant mass of 1.8--2 TeV will entail a scale $\fR \sim$ 6--8 TeV, leading to a negligible parameter $\bar{\epsilon}\sim 10^{-4}$ and suppressing therefore all the linear and higher $\bar{\epsilon}$--effects induced by the RH and LRH operators. These effects could be enhanced either via larger strength of the coefficient $c_{C,LR}$, or via NP effects from the right handed gauge sector around the EW scale $\fL$ together with a strength contribution of $c_{C,LR}$ around its maximal bound. It will be seen that the former scenario spoils the EW gauge masses, whereas the latter one points towards $\bar{\epsilon}\sim 10^{-2}$. The set of relevant non-linear LH, RH, and LRH operators will be completely identified for the latter $\bar{\epsilon}$--regimes and by disregarding: i) irrelevant LH operators with negligible physical impact on the observables considered for the $hVV$--bounds, ii) irrelevant operators for the non-linear realization of the dynamics and redundant for the massless fermion case; iii) operators without any direct contribution to the pure gauge  and gauge--Higgs couplings. It will be shown also that the diboson excess suggests to parametrize low energy effects via a much smaller effective operator basis as the $\bar{\epsilon}$--suppression entails. 

This work is split into: Sect.~\ref{EffectiveLagrangian} describes the EW effective Lagrangian following the light dynamical Higgs picture in~\cite{Alonso:2012px,Brivio:2013pma,Gavela:2014vra,Alonso:2012pz,Yepes:2015zoa} (see also Ref.~\cite{Buchalla:2013rka,Buchalla:2012qq,Buchalla:2013eza}, and for a short summary on the subject~\cite{Brivio:2015hua}), focused only in the CP--conserving bosonic operators\footnote{See ~\cite{Cvetic:1988ey,Alonso:2012jc,Alonso:2012pz,Buchalla:2013rka} for non--linear analysis including fermions.}, and providing all the LH, RH and LRH operators of the model up to $p^4$--order in the effective expansion. The mixing effects for the gauge masses triggered by the LRH operators and the corresponding gauge physical masses are analysed in~\ref{Rotating-mass-matrices}. The effects by integrating out the RH fields are studied in~\ref{Integrating-out}. Sect.~\ref{pheno} describes the phenomenology implied by the model and the allowed ranges for the involved coefficients weighting the effective operators. Sect.~\ref{Diboson-excess} comments on the recently observed diboson excess and its implications for the operator basis of our model, in particular on the relevant LH, RH and LRH operators for each one of the situations dictated by $\bar{\epsilon}$. Finally, Sect.~\ref{Conclusions} summarizes the main results.


\section{Effective Lagrangian}
\label{EffectiveLagrangian}

\nt The underlying strong dynamics assumed for this framework entails effective NP departures with respect to the SM Lagrangian $\LL_0$ and will be encoded through $\LL_\text{chiral}$ as
\be
\begin{aligned}
\LL_\text{chiral} = \LL_0\,+\,\LL_{0,R}
\,+\,\Delta \LL_{\text{CP}}\,+\,\Delta\LL_{\text{CP},LR}\,.\\[-4mm]
\label{Lchiral}
\end{aligned}
\ee 

\nt Focusing only on the bosonic interacting sector, the first three pieces in $\LL_\text{chiral}$ reads as
\be
\begin{aligned}
\LL_0 =& -\dfrac{1}{4}\,\BBd\,\BBu\,\,-\,\,\dfrac{1}{4}\,W^a_{\mu\nu,\,L}\,W^{\mu\nu,\,a}_L\,\,-\,\,
\dfrac{1}{4}\,G^a_{\mu\nu}\,G^{\mu\nu,\,a}\,\,+\,\,\\ \\
&\,\,+\,\,\frac{1}{2} (\derp_\mu h)(\derp^\mu h) - V (h)-\dfrac{\fL^2}{4}\Tr\Big(\VLLmuu\VLLmud\Big)\left(1+\frac{h}{\fL}\right)^2\,, \\ \\ 
\end{aligned}
\label{LLO}
\ee
\beq
\begin{aligned}
\LL_{0,R}= -\dfrac{1}{4}\,W^a_{\mu\nu,\,R}\,W^{\mu\nu,\,a}_R\,-\,\frac{\fR^2}{4}\,\Tr\Big(\VLRmuu\,\VLRmud\Big)\left(1+\frac{h}{\fL}\right)^2\,,\\ 
\label{LLO-Right}
\end{aligned}
\eeq
\nt where the adjoints $SU(2)_{L(R)}$--covariant vectorial $\VLLRmuu$, together with the covariant scalar $\TLLR$ objects, are defined as
\be
\VLLRmuu \equiv \left(\DL^\mu \UHLR\right)\,\UHLR^\dagger\,, \qquad\qquad
\TLLR \equiv \UHLR\,\tau_3\,\UHLR^\dagger\,,
\label{EFT-building-blocks}
\ee
\nt with the corresponding covariant derivative for both of the Goldstone matrices $\UHLR(x)$ introduced as
\be
\DL^\mu \UHLR \equiv \derp^\mu \UHLR \, + \,\frac{i}{2}\,\gLR\,W^{\mu,a}_{L(R)}\,\tau^a\,\UHLR - \frac{i}{2}\,g'\,B^\mu\, \UHLR\,\tau^3 \,,
\label{Covariant-derivatives}
\ee
\nt with the $SU(2)_L$, $SU(2)_R$ and $U(1)_{B-L}$ gauge fields denoted by $W^{a\mu}_L$, $W^{a\mu}_R$ and $B^\mu$ correspondingly, and the associated gauge couplings $\gL$, $\gR$ and $g'$ respectively. The usual SM strength gauge kinetic terms canonically normalized, the $h$--kinetic terms and the effective scalar potential $V(h)$ are present at $\LL_0$ in~\eqref{LLO}. The $W^\pm_L$ and $Z_L$ masses (before considering the corresponding RH and mixed LRH handed terms introduced a posteriori) and their couplings to the Higgs field $h$ can be read from the last term in the second line of $\LL_0$. The custodial breaking $p^2$--operator turns out to be strongly bounded phenomenologically, being thus left for the NP departures analysed later~\cite{Brivio:2013pma}.  The scale factor of $\Tr\left(\VLLmuu\,\VLLmud\right)$ entails GB--kinetic terms canonically normalized, in agreement with the $\UHL$--definition in~\eqref{Goldstone-matrices}. The corresponding $SU(2)_R$--counterparts for the strength gauge kinetic term and the custodial conserving operator are encoded at $\LL_{0,R}$ in~\eqref{LLO-Right}, implying thus an additional scale $\fR$ that encodes the new high energy scale effects introduced in the scenario once the SM local symmetry group $\cGSM$ is extended to $\cG$.  

Operators mixing the LH and RH-covariant are also constructable in this approach via the proper insertions of the Goldstone matrices $\UHL$ and $\UHR$, more specifically, through the following definitions~\cite{Yepes:2015zoa}
\be
\begin{aligned}
\widetilde{\VL}^\mu_\chi &\equiv \UH^\dagger_\chi\,\VLchimuu\,\UH_\chi,\, 
\qquad 
\widetilde{\TL}_\chi \equiv \UH^\dagger_\chi\,\TLchi\,\UH_\chi,\, \qquad
\WWchiut \equiv \UH^\dagger_\chi\,\WWchiu\,\UH_\chi\,,
\label{Vtilde-Ttilde-Wtilde}
\end{aligned}
\ee

\nt where $\WWchiu\equiv W^{\mu\nu,a}_\chi\tau^a/2$ ($\chi=L,R$) (see Appendix~\ref{App:Operators-list-LR}). Non--zero NP departures with respect to those described in $\LL_0\,+\,\LL_{0,R}$ will be parametrized through the remaining last two pieces in $\LL_\text{chiral}$, i.e $\Delta\LL_{\text{CP}}$ and $\Delta\LL_{\text{CP},LR}$. The former encodes all those effective non--linear operators made out of purely LH or RH covariant objects up to the $p^4$--order~\cite{Yepes:2015zoa}, being split as
\be
\Delta \LL_{\text{CP}}=\Delta \LL_{\text{CP},L}+\Delta \LL_{\text{CP},R}
\label{DeltaL-CP-even}
\ee

\nt where the suffix $L(R)$ labels all those operators set constructed out by means of the $SU(2)_{L(R)}$ building blocks in~\eqref{Vtilde-Ttilde-Wtilde}. The contribution $\Delta \LL_{\text{CP},L}$ has already been provided in~\cite{Alonso:2012px,Brivio:2013pma} in the context of purely EW chiral effective theories coupled to a light Higgs, whereas part of $\Delta \LL_{\text{CP},L}$ and $\Delta \LL_{\text{CP},R}$ were partially analysed for the left--right symmetric frameworks in~\cite{Zhang:2007xy,Wang:2008nk}, and finally completed in the recent work~\cite{Yepes:2015zoa} (for the corresponding CP--violating part see~\cite{Yepes:2015qwa}). Both of the contributions $\Delta \LL_{\text{CP},L}$ and $\Delta \LL_{\text{CP},R}$ can be further written down as 
\beq
\hspace*{-1mm}
\Delta \LL_{\text{CP},L}=c_G\cP_G(h)+c_B\cP_B(h)+\sum_{i=\{W,C,T\}}c_{i,L}\cP_{i,L}(h)\,\,+\sum_{i=1}^{26} c_{i,L}\cP_{i,L}(h) \,+\, c_H \cP_H(h) \,+\, c_{\Box H} \cP_{\Box H}(h)
\label{DeltaL-CP-even-L}
\eeq

\be
\Delta\LL_{\text{CP},R}= \sum_{i=\{W,C,T\}}c_{i,R}\cP_{i,R}(h)\,\, \,+\,\,\,\sum_{i=1}^{26} c_{i,R}\cP_{i,R}(h)
\label{DeltaL-CP-even-R}
\ee

\nt where $c_{B}$, $c_{G}$ and  $c_{i,\chi}$ are model--dependent constant coefficients, whilst the first three terms of $\Delta \LL_{\text{CP},L}$ in~\eqref{DeltaL-CP-even-L} and the first term in~\eqref{DeltaL-CP-even-R} can be jointly written as
\beq
\begin{aligned}
\cP_G(h)\,\,  &= -\frac{g^2_s}{4}\,G_{\mu\nu}^a\,G^{\mu\nu}_a\,\cF_G(h) \\
\cP_B(h)\,\,  &= -\frac{g'^2}{4}\,B_{\mu\nu}\,B^{\mu\nu}\,\cF_B(h) \\ 
\cP_{W,\,\chi}(h)\,\,  &= -\frac{\gchi^2}{4}\, W_{\mu\nu,\,\chi}^a\,W^{\mu\nu,\,a}_\chi\,\cF_{W,\,\chi}(h)  \\  
\cP_{C,\,\chi}(h)\,\,  &= - \frac{\fchi^2}{4}\Tr\Big(\VLchimuu\,\VLchimud\Big) \,\cF_{C,\,\chi}(h) \\ 
\cP_{T,\,\chi}(h)\,\,  &= \frac{\fchi^2}{4}\, \Big(\Tr\Big(\TLchi\,\VLchimuu\Big)\Big)^2\,\cF_{T,\,\chi}(h)  
\label{GT}
\end{aligned}
\eeq    
\nt with suffix $\chi$ labelling again as $\chi=L,R$, and the generic $\cF_i(h)$--function of the scalar singlet $h$ is defined for all the operators following definition~\eqref{F}. Finally, the last two terms in the second line of $\Delta \LL_{\text{CP},L}$ in~\eqref{DeltaL-CP-even-L} account for all the possible pure Higgs interactions, with the $p^2$ and $p^4$--operators $\cP_H$ and $\cP_{\Box H}$ respectively as
\be
\cP_H(h) = \frac{1}{2}\,(\derp_\mu h)^2\,\cF_H(h)\,,
\qquad
\cP_{\Box H}=\frac{1}{v^2}\,(\Box h)^2\,\cF_{\Box H}(h)\,.
\label{Opxih}
\ee

\nt The set of 26 CP--conserving pure gauge and gauge--$h$ non--linear operators encoded by $\cP_{i,\,L}(h)$ (fourth term in $\Delta \LL_{\text{CP},\,L}$, Eq.~\eqref{DeltaL-CP-even-L}) have completely been listed in~\cite{Alonso:2012px,Brivio:2013pma}. On the other hand, the symmetric counterpart made out of the 26 CP-conserving operators $\cP_{i,\,R}(h)$ (second term in $\Delta \LL_{\text{CP},\,R}$, Eq.~\eqref{DeltaL-CP-even-R}) were recently reported in~\cite{Yepes:2015zoa}. In total there are 52 non--linear operators, 38 of them (19 $\cP_{i,\,L}$ + 19 $\cP_{i,\,R}$) had already been listed in~\cite{Zhang:2007xy,Wang:2008nk}, whilst 14 additional operators were found (7 $\cP_{i,\,L}$ + 7 $\cP_{i,\,R}$) in~\cite{Yepes:2015zoa} with respect to~\cite{Zhang:2007xy,Wang:2008nk} (seven of them, corresponding to $\chi = L$, were already reported in~\cite{Alonso:2012px,Brivio:2013pma}). See~\cite{Yepes:2015zoa} for the complete list of operators $\Delta \LL_{\text{CP},\,L}$ and related discussion.

Finally, $\Delta\LL_{\text{CP},LR}$ parametrizes any possible mixing interacting term between the $SU(2)_L$ and $SU(2)_R$--covariant objects up to the $p^4$--order in the Lagrangian expansion, permitted by the underlying left--right symmetry, and encoded through~\cite{Yepes:2015zoa}
\be
\Delta\LL_{\text{CP},LR}=\sum_{i=\{W,C,T\}}c_{i,LR}\cP_{i,LR}(h)\,\, \,+\,\,\,\sum_{i=2,\,i\neq 4}^{26} c_{i(j),LR}\,\cP_{i(j),LR}(h)\,,
\label{DeltaL-CP-even-LR}
\ee

\nt where the index $j$ spans over all the possible operators that can be built up from the set of 26 non--linear operators $\cP_{i,\chi}$ in~\eqref{DeltaL-CP-even-L}--\eqref{DeltaL-CP-even-R} (fourth and second terms respectively), and here labelled as $\cP_{i(j),LR}$ (as well as their corresponding coefficients $c_{i(j),LR}$). The first term in $\Delta\LL_{\text{CP},LR}$ encodes the non-linear mixing operators
\be
\begin{aligned}
\cP_{W,\,LR}(h)\,\,  &= -\frac{1}{2}\,\gL\,\gR\,\Tr\left(\WWLut\,\WWRdt\right)\,\cF_{W,\,LR}(h)\,,  \\ \\
\cP_{C,\,LR}(h)\,\,  &=  \frac{1}{2}\,\fL\,\fR\,\Tr\Big(\VLLmuut\VLRmudt\Big)\,\cF_{C,\,LR}(h)\,, \\ \\
\cP_{T,\,LR}(h)\,\,  &= \frac{1}{2}\,\fL\,\fR\,\Tr\Big(\TLLt\,\VLLmuut\Big)\,\Tr\Big(\TLRt\,\VLRmudt\Big)\cF_{T,\,LR}(h)\,,  
\label{WT-LR}
\end{aligned}
\ee

\nt corresponding to the ``mixed" versions of $\{\cP_{W,\chi},\,\cP_{C,\chi},\,\cP_{T,\chi}\}$ in~\eqref{GT}, with $\cP_{W,\,LR}$ missing in~\cite{Zhang:2007xy,Wang:2008nk}. The complete set of operators $\cP_{i(j),LR}$ in the second term of $\Delta\LL_{\text{CP},LR}$ have been fully and recently listed in the reference~\cite{Yepes:2015zoa}. Both of the previously described CP--conserving contributions $\Delta\LL_{\text{CP}}$ and $\Delta\LL_{\text{CP},LR}$ have been completely listed and studied in~\cite{Yepes:2015zoa}, whereas their corresponding CP--violating counterparts analysed in~\cite{Yepes:2015qwa}. Notice that in the unitary gauge, non-zero mass mixing terms among the LH and RH gauge fields are triggered by the operator $\cP_{C,\,LR}$, leading to diagonalize the gauge sector in order to obtain the required physical gauge masses.

\nt Non--linear approaches have already been linked to the linear effective scenarios explicitly implementing the SM Higgs doublet through~\cite{Alonso:2012px,Brivio:2013pma}, where all the operators in~\eqref{DeltaL-CP-even-L} were respectively weighted by powers of $\xi\equiv v^2/\fL^2$, in order to keep track of their corresponding operator siblings in the linear regime. In fact, operators in~\eqref{LLO}, and those in the first line of~\eqref{DeltaL-CP-even-L}, as well as $\cP_{1-5,L}$, had been already 
pointed out in the analysis of the linear--non linear connection of the SILH framework~\cite{Giudice:2007fh,Buchalla:2014eca}. Indeed, for the $\xi$--small limit, all the operators weighted by $\xi^{n\geqslant2}$ are negligible and the resulting Lagrangian is directly linked to the SILH treatment. Similar analysis has been recently done for the Higgs portal to scalar dark matter in the context of non-linearly realised electroweak symmetry breaking~\cite{Brivio:2015kia}. For the assumed non--linear scenario in this work, such linking between both of the EFT sides implies the corresponding left--right symmetric extension of the effective linear approaches and it is beyond the scope of this work.

Furthermore, the transformation properties under the parity symmetry $P_{LR}$ of the mixing operators from $\Delta\LL_{\text{CP},LR}$ in~\eqref{DeltaL-CP-even-LR} were analysed in~\cite{Yepes:2015zoa}. In fact it was explicitly shown that a set of $p^4$--operators, more specifically, $\cP_{18(3-6),LR}(h)$, triggers the breaking of $P_{LR}$, alike in the context of a general effective $SO(5)/SO(4)$ composite Higgs model scenario~\cite{Contino:2011np}, where $P_{LR}$ was shown to be an accidental symmetry up to $p^2$--order and broken by several $p^4$--operators~\cite{Contino:2011np,Alonso:2014wta}. 

An interesting feature of this scenario arises once the gauge field content $W_R$ is integrated out from the physical spectrum at low energies. In fact, the RH terms encoded through the Lagrangian $\LL_{0,R}$ in~\eqref{LLO-Right}, will impact directly onto the left handed ones of $\LL_0$ in~\eqref{LLO}. Similarly, the RH and LRH terms parametrized by $\Delta\LL_{\text{CP},R}$ and $\Delta\LL_{\text{CP},LR}$ in~ \eqref{DeltaL-CP-even-R} and~\eqref{DeltaL-CP-even-LR} will affect those ones from $\Delta \LL_{\text{CP},L}$ in~\eqref{DeltaL-CP-even-L}. These modifications will alter therefore the effective couplings sourced by the whole Lagrangian $\LL_0\,+\,\Delta \LL_{\text{CP},L}$, specifically, the triple and quartic gauge boson couplings, as well as the gauge bosons--Higgs couplings. Non--zero contributions on the EWPD parameters $S$ and $T$ will also be induced as it will be seen later. At higher energies, when the right handed gauge fields $W^\mu_R$ are still playing a role, mixing effects among them and the left handed gauge field sector will be triggered by the  $\cP_{C,LR}$ in~\eqref{WT-LR}. Those effects are treated in the following section.


\subsection{Rotating to the physical sector}
\label{Rotating-mass-matrices}

\nt At the unitary gauge, the non-linear chiral effective Lagrangian $\LL_\text{chiral}$ in ~\eqref{Lchiral} sources the following mass terms for both of the charged and neutral gauge sectors
\be
\LL_\text{chiral}\,\supset \,\widehat{\cW}_{\mu }^{+T}\,\cM_{\cW}\,\widehat{\cW}^{-\mu }\,\,+\,\,\frac{1}{2}\,\widehat{\cN}_{\mu }^T\,\cM_{\cN}\,\widehat{\cN}^{\mu }
\label{Mass-Lagrangian}
\ee

\nt with the gauge basis defined by
\be
\widehat{\cW}_{\mu }^\pm\equiv 
\left(
\begin{array}{c}
 W_{\mu,L}^\pm \\ [3mm]
 W_{\mu,R}^\pm \\
\end{array}
\right)\,, \qquad\qquad 
\widehat{\cN}_{\mu }\equiv \left(
\begin{array}{c}
 W_{\mu ,L}^3 \\ [3mm]
 W_{\mu ,R}^3 \\[3mm]
 B_{\mu } \\
\end{array}
\right)\,,
\label{Gauge-basis}
\ee

\nt where the charged fields $W_{\mu,\chi}^\pm$ are defined as usual
\be
W_{\mu,\chi}^\pm\equiv \frac{W_{\mu,\chi}^1\,\mp\,i\,W_{\mu,\chi}^2}{\sqrt{2}}\,,\qquad\qquad \chi=L,\,R\,,
\label{Charged-fields}
\ee

\nt and the mass matrix for the charged sector written as
\be
\cM_{\cW}=
 \frac{1}{4}\,\gL^2\,\fL^2\,\left(
\begin{array}{cc}
1+c_{C,\,L} & -\frac{1}{\sqrt{\lambda }} \,c_{C,\,LR} \\ \\
- \frac{1}{\sqrt{\lambda }}\,c_{C,\,LR} & \frac{1}{\lambda }\left(1+c_{C,\,R}\right) \\
\end{array}\right)
\label{Charged-gauge-matrix}
\ee
\nt with the parameter $\lambda$ defined as
\be
\lambda\equiv\frac{\gL^2}{\gR^2}\,\epsilon^2,\,\qquad\qquad \epsilon\equiv\frac{\fL}{\fR}\,. 
\label{xi}
\ee

\nt The corresponding mass matrix for the neutral sector reads as
\be
\small{
\begin{aligned}
&\cM_{\cN}=\\ \\
&\hspace*{-0.5cm}\frac{g_L^2 \mathit{f}_L^2}{4}\left(
\begin{array}{ccc}
1+\alpha_L & -\frac{\alpha _{LR}}{\sqrt{\lambda}} & -\frac{g'}{\gL}
\left(1+\alpha _L-\frac{\fR}{\fL}\alpha _{LR}\right) \\ \\
 -\frac{\alpha _{LR}}{\sqrt{\lambda }} & \frac{1+\alpha _R}{\lambda } & \frac{g'}{\gL\sqrt{\lambda }}\left(\alpha _{LR}-\frac{\fR}{\fL}\left(1+\alpha _R\right)\right) \\ \\
-\frac{g'}{\gL}
\left(1+\alpha _L-\frac{\fR}{\fL}\alpha _{LR}\right) & \frac{g'}{\gL\sqrt{\lambda }}\left(\alpha _{LR}-\frac{\fR}{\fL} \left(1+\alpha _R\right)\right) & \frac{g'^2 }{\gL^2 }
\left(1+\alpha _L-\frac{2\fR}{\fL}\alpha _{LR} +\frac{\fR^2}{\fL^2}\left(1+\alpha _R\right)\right) \\ \\
\end{array}
\right)
\label{Neutral-gauge-matrix}
\end{aligned}
}
\ee

\nt where it have been introduced the definitions
\be
\alpha _{LR} \equiv c_{C,\,LR}+2\,c_{T,\,LR}\,,\qquad\qquad
\alpha _\chi \equiv c_{C,\,\chi}-2\,c_{T,\,\chi}\,,\qquad \qquad \chi=L,\,R\,.
\label{alphas}
\ee

\nt The latter matrices can be diagonalized via the following field transformations 
\be
\widehat{\cW}_{\mu }^\pm\equiv 
\cR_{\cW}\,\cW_{\mu }^\pm\,, \qquad\qquad \widehat{\cN}_{\mu }\equiv \cR_{\cN}\,\cN_{\mu }\,,
\label{Gauge-field-transformations}
\ee

\nt with the mass eigenstate basis defined by 
\be
\cW_{\mu }^\pm\equiv 
\left(
\begin{array}{c}
W_{\mu}^\pm \\[3mm]
W^{\prime\pm}_{\mu} 
\end{array}
\right)\,, \qquad\qquad 
\cN_{\mu }\equiv \left(
\begin{array}{c}
A_{\mu} \\[3mm]
Z_{\mu} \\[3mm]
Z^\prime_{\mu} \\
\end{array}
\right)\,.
\label{Mass-basis}
\ee

\nt The rotation matrix for the charged sector in Eq.~\eqref{Gauge-field-transformations} is given by

\be	
\cR_{\cW}=
\left(
\begin{array}{cc}
 \cos\zeta & -\sin\zeta \\ \\
 \sin\zeta & \cos\zeta \\
\end{array}
\right)\,,\qquad\qquad \tan\zeta= -\frac{\sqrt{\lambda }}{1-\lambda}\, c_{C,\,LR}\,, 
\label{Charged-rotation}
\ee

\nt with the mixing angle $\zeta$ for the charged sector directly depending on the parameter $\lambda$ and the mixing coefficient $c_{C,\,LR}$. Expanding up to the order $\cO(\lambda^2)$ in the limit $\fL\ll\fR$, the charged gauge masses are
\be
M^2_W \,\,\simeq\,\, \frac{1}{4}\,\gL^2\,\fL^2\,\Big(1+c_{C,L}+c_{C,R} - \lambda\,c^2_{C,LR}\Big)\,,\qquad  M^2_{W^\prime} \,\,\simeq\,\, \frac{1}{4}\,\gR^2\,\fR^2\,\Big(1+c_{C,R} + \lambda\,c^2_{C,LR}\Big)\,. \\
\label{Charged-masses-expanded}
\ee

\nt For the the neutral sector we have the real symmetric 3$\times$3 matrix in~\eqref{Neutral-gauge-matrix}, then an orthogonal rotation via the Euler-type angles parametrization is in order to diagonalize it. Such angles turn out to be the Weinberg mixing angle $\theta_W$ and the analogous mixing angle $\theta_R$ for the $SU(2)_R\otimes U(1)_{B-L}$ sector defined correspondingly as
\be
\cos\theta_W\equiv c_W = \frac{\gL}{\sqrt{\gL^2 + g^2_Y}},\,\qquad \sin\theta_W\equiv s_W = \frac{g_Y}{\sqrt{\gL^2 + g^2_Y}}\,,
\label{SM-mixing-angle}
\ee

\be
\cos\theta_R\equiv c_R = \frac{\gR}{\sqrt{\gR^2 + g'^2}}=\frac{g_Y}{g'},\,\qquad \sin\theta_R\equiv s_R = \frac{g'}{\sqrt{\gR^2 + g'^2}}=\frac{g_Y}{\gR}
\label{Right-B-L-mixing-angle}
\ee

\nt where it has been employed in~\eqref{Right-B-L-mixing-angle} the link among the $SU(2)_L$, $U(1)_{B-L}$ and the SM hypercharge gauge couplings as
\be
\frac{1}{g_R^2} + \frac{1}{g'^2}= \frac{1}{g_Y^2}\,.
\label{B-L-gauge-coupling}
\ee

\nt The third angle $\phi$ can be linked to the latter two through
\be
\tan\phi \,\,\simeq\,\,\epsilon ^2 \frac{g_L}{g_R}\frac{s^2_R c_R}{c_W} + \cO(\epsilon^4)\,.
\label{Third-mixing-angle}
\ee

\nt The rotation matrix for the neutral sector becomes parametrized then as
\be
\cR_{\cN}=
\left(
\begin{array}{ccc}
 s_W & c_W  & - \frac{g_L}{g_R}\,\epsilon^2\,s^2_R\,c_R
\\ \\
 c_W s_R & - s_R s_W & c_R  \\ \\
 c_R c_W & -c_R s_W  & -  s_R \\
\end{array}
\right)\,.
\label{Neutral-rotation}
\ee

\nt Expanding up to the order $\cO(\lambda^2)$ (in the limit $\fL\ll\fR$) the neutral gauge masses are

\be
M^2_Z\,\,\simeq\,\,\frac{M^2_W}{c^2_W}\left(1+\alpha_L\right)\,, \qquad 
M^2_{Z^\prime}\,\,\simeq\,\,\frac{M^2_{W'}}{c^2_R}\left(1+\alpha_R-2\,\alpha _{LR}\,s^2_R\,\epsilon\right)\,
\label{Neutral-masses-expanded}
\ee

\nt with $\tan\theta_R\equiv t_R$. The well measured $M_Z$--mass strongly constrains the coefficient $\alpha_L$ in~\eqref{Neutral-masses-expanded}, and therefore the contribution from the operators $\cP_{C,\,L}(h)$ and $\cP_{T,\,L}(h)$. Likewise, the $M_W$--mass bounds tightly constrains the contribution from $\cP_{C,\,R}(h)$ in~\eqref{Charged-masses-expanded}. A mass prediction for the extra neutral gauge field $Z'$ can be inferred from~\eqref{Neutral-masses-expanded} in terms of the $W'$--mass and the RH gauge coupling $\gR$ via the mixing angle $\theta_R$ in~\eqref{Right-B-L-mixing-angle}. In fact, interpreting the observed excess at the ATLAS and CMS Collaborations around invariant mass  of 1.8--2 TeV to be induced by a
$W'$--contribution, and assuming the coupling $\gR$ in the range $\gR \approx 0.45-0.6$ as determined in~\cite{Dobrescu:2015qna} by comparing the $W^\prime$ production cross section to the CMS dijet excess \cite{Khachatryan:2015sja}, it is possible to predict the mass range
$2.4\,\text{TeV}\,<\,M_{Z^\prime}\,<\,4\,\text{TeV} $. A more detailed interpretation of the diboson excess via a left--right non-linear Higgs approach can be found in~\cite{Yepes-III}.

A higher energy scale $\fR$ points in general towards higher masses $M_{W^\prime}$ and $M_{Z^\prime}$, additionally entailing  a vanishing mixing angle $\zeta$ among the charged gauge fields $W_{\mu,L}^\pm$ and $ W_{\mu,R}^\pm$ as $\lambda \rightarrow 0$ (see Eqs.~\eqref{Gauge-basis},~\eqref{Mass-basis} and~\eqref{Charged-rotation}), neither a mixing among the set of neutral fields $\{W_{\mu ,L}^3,\,B_{\mu}\}$ with the field  $W_{\mu ,R}^3$ (see Eqs.~\eqref{Gauge-basis},~\eqref{Mass-basis} and~\eqref{Neutral-rotation}) and therefore right handed gauge fields directly linked to the eigenstate basis as $ W_{\mu,R}^\pm = W^{'\pm}_{\mu}$ and $W_{\mu ,R}^3 = - Z^\prime_{\mu}$. Heavy right handed gauge fields can thus be integrated out from the physical spectrum of the model, triggering therefore physical effects that will be manifested at lower energies in the effective Lagrangian. Such procedure together with the induced effects it leads to, will be analysed via equations of motion for the right handed fields in the following section.

\subsection{Integrating-out heavy right handed fields}
\label{Integrating-out}

\nt From the equations of motion for the gauge and Higgs fields (Eqs.~\eqref{W-EOM}-\eqref{h-EOM}, Appendix~\ref{App:EOM}) it is possible to integrate out the RH gauge fields from the physical spectrum. In fact, at low energies one obtains from the EOM
\be
\VLRmuu\,\equiv\,-\bar{\epsilon}\,\,\VLLmuu\,,\qquad
\text{with}\qquad\bar{\epsilon}\equiv \epsilon\,c_{C,LR}
\label{Gauge-field-EOM}
\ee

\nt that can be translated into the unitary gauge as
\be
W_{\mu,\,R}^\pm \quad\Rightarrow\quad -\frac{\gL}{\gR}\,\bar{\epsilon}\,\,W_{\mu,\,L}^\pm\,,\qquad  W_{\mu ,R}^3\quad\Rightarrow\quad \frac{g'}{\gR}\left(1+\bar{\epsilon}\right)B_{\mu } -\frac{\gL}{\gR}\,\bar{\epsilon}\,\,W_{\mu ,L}^3
\label{Gauge-field-EOM-unitary-gauge}
\ee

\nt After such field redefinition, all the couplings and operator coefficients will be shifted, affecting thus the final form for the TGV couplings, anomalous quartic gauge and gauge-Higgs couplings, and modifying as well the final expressions for the EWPD parameters as it will be seen in the next sections. The final Lagrangian at low energies, here denoted by $\bar{\LL}_\text{chiral}$ with respect to $\LL_\text{chiral}$ in~\eqref{Lchiral}, will be given by
\be
\begin{aligned}
\bar{\LL}_\text{chiral} = \bar{\LL}_0\,+\,\Delta \bar{\LL}_{\text{CP}}\,,\\[-5mm]
\label{Lchiral-redefined}
\end{aligned}
\ee 

\nt where the first component reads as
\be
\hspace*{-0.3cm}
\begin{aligned}
&\bar{\LL}_0 =\\ \\
& -\dfrac{1}{4}\,(1+\alpha_B)\,\BBd\,\BBu\,\,-\,\,
\dfrac{1}{2}\,\alpha_{WB}\,B_{\mu\nu}\,\Tr\Big(\TL\,\WWu\Big)
\,\,-\,\,\dfrac{1}{4}\,(1+\alpha_W)\,W^a_{\mu\nu}\,W^{\mu\nu,\,a}\,\,-\,\,
\dfrac{1}{4}\,G^a_{\mu\nu}\,G^{\mu\nu,\,a}\,\,+\,\,\\ \\
&\,\,+\,\,\frac{1}{2} (\derp_\mu h)(\derp^\mu h) - V (h)-\dfrac{f^2}{4}\Tr\Big(\VLmuu\,\VLmud\Big)\left(1+\frac{h}{f}\right)^2\,. \\
\end{aligned}
\label{LLO-redefined}
\ee

\nt Hereafter the fields $W^{\mu\nu,\,a}_L$ and $\VLLmuu$ are properly relabelled as $W^{\mu\nu,\,a}_L \rightarrow W^{\mu\nu,\,a}$ and $\VLLmuu\,\rightarrow\,\VLmuu$ (with $\gL \rightarrow g$, $g$ as  $SU(2)_Y$--gauge coupling). The shifts in the strength gauge kinetic terms are encoded through the coefficients $\alpha_B$, $\alpha_{WB}$ and $\alpha_W$, and defined by
\be
\begin{aligned}
\alpha_B\,\equiv \frac{g'^2}{\gR^2}\left(1+\bar{\epsilon}\right)^2,\,\qquad
\alpha_{WB}\,\equiv \frac{g'}{2 \gR}\left(1-2\,\frac{\gL}{\gR}\,\bar{\epsilon}\right)\left(1+\bar{\epsilon}\right),\, 
\qquad 
\alpha_W\,\equiv \,-\frac{\gL}{\gR}\,\bar{\epsilon}\\ 
\end{aligned}
\label{Strength-kinetic-shifts}
\ee

\nt and the scale $f$ is given by the redefinition
\be
f\,\equiv\,\fL\,\sqrt{1+c^2_{C,\,LR}}\,.
\label{Redefined-fL}
\ee

\nt The EW gauge mass $M_W$ strongly constrains the quadratic contribution of the LRH operator $\cP_{C,LR}(h)$ in~\eqref{Redefined-fL} as 
\be
-0.02<c_{C,\,LR}<0.02\,.
\label{cLR-bound}
\ee

\nt Notice that a mixing term in the kinetic gauge sector is induced at low energies, driving thus additional effects when diagonalizing such sector, as well as a non-zero contribution to the $S$--parameter as it will be seen a posteriori. The second component in~\eqref{Lchiral-redefined} is basically the Lagrangian in $\eqref{DeltaL-CP-even-L}$ but the gluonic operator $\cP_G(h)$, and with the coefficients $\{c_{B},\,c_{i,\chi}\}$ properly redefined as $\{c_{B},\,c_{i,\chi}\} \rightarrow \{\tilde{c}_{B},\,\tilde{c}_{i,\chi}\}$ in order to account for the induced effects after removing away the RH gauge fields from the physical spectrum. Table~\ref{Redefined-coefficients} displays all the initial contribution $c_{i,L}$ from the LH operators $\mathcal{P}_{i,L}$ (1st column) receiving a contribution from the RH $\mathcal{P}_{i,R}$ (2nd column), plus a combination from the mixing LRH $\mathcal{P}_{\text{i(j)}}$ (3rd column), and for each one of the LH non-linear operators $\mathcal{P}_{i,L}$ (indicated at the 4th column). The sum of the values at the first, second and third columns determines the coefficients $\tilde{c}_{i,L}$. It is possible to infer that for the limiting hierarchical case $\fL\ll \fR$ at low energies, the set of non-linear operators 
\be
\{\cP_B,\,\cP_{C,L},\,\cP_{T,L},\,\cP_{1,L},\,\cP_{2,L},\,\cP_{4,L}\}\,
\label{Sensitive-operators}
\ee

\nt is sensitive to the contributions from both of the RH operators

\be
\{\cP_{C,R},\,\cP_{T,R},\,\cP_{W,R},\,\cP_{1,R},\,\cP_{12,R}\}\,
\label{Right-operators}
\ee

\nt and the mixing LRH set
\be
\{\cP_{C,LR},\,\cP_{T,LR},\,\cP_{W,LR},\,\cP_{\text{3(2)}},\,\cP_{\text{12(1)}},\,\cP_{\text{13(2)}},\,\cP_{\text{17(2)}}\}\,.
\label{Left-Right-operators}
\ee

\begin{table}
\hspace*{-2.2cm}
\centering
\small{
\renewcommand{\arraystretch}{0.0}
\begin{tabular}{c||cc||c}
\hline\hline
$\bf \cP_{i,L}$  &  $\bf \cP_{i,R}$  &  $\bf\cP_{i(j),LR}$  &  $\bf i$\\
\hline\hline 
$ c_{i} $ &  
$\begin{array}{l}  
c_{W,R}-4 c_{1,R}-4 c_{12,R}+\bar{\epsilon}^2\,\left(c_{W,R}-4 c_{12,R}\right)+\\
\\[2mm]
+\,\bar{\epsilon}\left(2 c_{W,R}-4 c_{1,R}-8 c_{12,R}\right)
\end{array}$ &  $ - $ &  $ B $ \\[7mm]  
 $ c_{i,L} $ &  $ \bar{\epsilon}^2\,c_{i,R} $ &  $ -\bar{\epsilon}\,c_{i,LR} $ &  $ W $ \\[5mm]  
$ c_{i,L} $ &  $ c_{i,R} $ &  $ \left\{-2\, c_{C,LR},-2\,c_{T,LR}\right\} $ &  $ \{C,\,T\} $ \\[5mm]  
$ c_{i,L} $ &  
$ \frac{\bar{\epsilon}}{2}\left(-2 c_{i,R}+c_{W,R}-4 c_{12,R}\right)+\frac{\bar{\epsilon}^2}{2}\left(c_{W,R}-4 c_{12,R}\right) $ &  $ \frac{\bar{\epsilon}}{4}\left(4 c_{\text{12(1)}}-c_{W,LR}\right)+\frac{1}{4} \left(4 c_{\text{12(1)}}-c_{W,LR}\right) $ &  $ 1 $ \\[7mm]  
$ c_{i,L} $ &  $ \frac{\bar{\epsilon}^2 }{2}\left(2 c_{i,R}+c_{3,R}+2 c_{13,R}\right)+\frac{\bar{\epsilon}^3}{2}\left(c_{3,R}+2 c_{13,R}\right) $ &   
$\begin{array}{l}  
\frac{1}{2} \left(2 c_{\text{13(2)}}+c_{\text{3(2)}}\right)-\frac{\bar{\epsilon}^2}{2}\left(2 c_{\text{13(4)}}+c_{\text{3(4)}}\right)+\\
\\[1mm]
-\frac{\bar{\epsilon}}{2}\left(2 \left(-c_{\text{13(2)}}+c_{\text{13(4)}}+c_{\text{2(1)}}\right)-c_{\text{3(2)}}+c_{\text{3(4)}}\right)
\end{array}$  &  $ 2 $ \\[12mm]  
$ c_{i,L} $ &  $ -\bar{\epsilon}^3\,c_{i,R} $ &  $ \bar{\epsilon}^2\left(c_{\text{i(1)}}+c_{\text{i(4)}}\right)-\bar{\epsilon}\left(c_{\text{i(2)}}+c_{\text{i(3)}}\right) $ &  $ \text{3, 13} $ \\[5mm]  
$ c_{i,L} $ &  $ -\frac{\bar{\epsilon}}{a_{i,L}}\left(a_{i,R} c_{i,R}+a_{17,R} c_{17,R}\right)-\bar{\epsilon}^2\,\frac{a_{17,R}}{a_{i,L}}\,c_{17,R} $ &  $ \bar{\epsilon}\frac{a_{\text{17(2)}}}{a_{i,L}}\,c_{\text{17(2)}} +\frac{a_{\text{17(2)}}}{a_{i,L}}\,c_{\text{17(2)}}  $ &  $ 4 $ \\[5mm]  
$ c_{i,L} $ &  $ \bar{\epsilon}^2\, \frac{a_{i,R}}{a_{i,L}}\,c_{i,R} $ &  $ -\bar{\epsilon}\frac{a_{\text{i(1)}}}{a_{i,L}}\,c_{\text{i(1)}} -\bar{\epsilon}\,\frac{a_{\text{i(2)}}}{a_{i,L}}\,c_{\text{i(2)}}  $ &  $ \text{5, 10, 17, 19} $ \\[5mm]
$ c_{i,L} $ &  $ \bar{\epsilon}^4\,c_{i,R} $ &  $ \frac{\fL^2 \left(c_{\text{i(1)}}+c_{\text{i(2)}}\right)}{\fR^2}-\bar{\epsilon}\,c_{\text{i(3)}} -\bar{\epsilon}^3c_{\text{i(4)}}  $ &  $ \text{6, 26} $ \\[5mm]
$ c_{i,L} $ &  $ \bar{\epsilon}^2\,\frac{a_{i,R}}{a_{i,L}}\,c_{i,R} $ &  $ -\bar{\epsilon}\frac{a_{\text{i(1)}}}{a_{i,L}}\,c_{\text{i(1)}}  $ &  $ \text{7, 25} $ \\[5mm]
$ c_{i,L} $ &  $ \bar{\epsilon}^2\,\frac{a_{i,R}^2}{a_{i,L}^2}\,c_{i,R} $ &  $ -\bar{\epsilon}\frac{a_{\text{i(1)}}^2}{ a_{i,L}^2}\,c_{\text{i(1)}}  $ &  $ \text{8, 20, 21, 22} $ \\[5mm]
$ c_{i,L} $ &  $ \bar{\epsilon}^2\,c_{i,R} $ &  $ -\bar{\epsilon}\,c_{\text{i(1)}}  $ &  $ \text{9, 12, 15} $ \\[6mm]  
$ c_{i,L} $ &  $ \bar{\epsilon}^4\,c_{i,R} $ &  $ \bar{\epsilon}^2\left(c_{\text{i(1)}}+c_{\text{i(2)}}+c_{\text{i(5)}}\right)-\bar{\epsilon}\,c_{\text{i(3)}} -\bar{\epsilon}^3\,c_{\text{i(4)}}  $ &  $ 11 $ \\[6mm]
$ c_{i,L} $ &  $ -\bar{\epsilon}^3\,c_{i,R} $ &  $ -\bar{\epsilon}\left(c_{\text{i(1)}}+c_{\text{i(3)}}+c_{\text{i(5)}}\right)
 +\bar{\epsilon}^2\left(c_{\text{i(2)}}+c_{\text{i(4)}}+c_{\text{i(6)}}\right) $ &  $ 14 $ \\[6mm]
$ c_{i,L} $ &  $ -\bar{\epsilon}^3\,c_{i,R} $ &  $ -\bar{\epsilon}\left(c_{\text{i(1)}}+c_{\text{i(4)}}+c_{\text{i(6)}}\right)
 +\bar{\epsilon}^2\left(c_{\text{i(2)}}+c_{\text{i(3)}}+c_{\text{i(5)}}\right) $ &  $ 16 $ \\[8mm]
$ c_{i,L} $ &  $ -\bar{\epsilon}^3\frac{a_{i,R}}{a_{i,L}}\,c_{i,R} $ &  
$\begin{array}{l}  
-\frac{\bar{\epsilon}}{a_{i,L}}\left(c_{\text{i(1)}} a_{\text{i(1)}}+c_{\text{i(3)}} a_{\text{i(3)}}+c_{\text{i(6)}} a_{\text{i(6)}}\right)+\\
\\[1mm]
+\frac{\bar{\epsilon}^2 }{\fR^2 a_{i,L}}\left(c_{\text{i(2)}} a_{\text{i(2)}}+c_{\text{i(4)}} a_{\text{i(4)}}+c_{\text{i(5)}} a_{\text{i(5)}}\right)
\end{array}$  &  $ 18 $ \\[13mm]
$ c_{i,L} $ &  $ \bar{\epsilon}^4\,c_{i,R} $ &  
$\begin{array}{l}  
-\bar{\epsilon}\left(c_{\text{i(3)}}+c_{\text{i(6)}}\right)-\bar{\epsilon}^3 \left(c_{\text{i(4)}}+c_{\text{i(5)}}\right)+\\
\\[2mm]
+\bar{\epsilon}^2\left(c_{\text{i(1)}}+c_{\text{i(2)}}+c_{\text{i(7)}}\right)
\end{array}$  &  $ 23 $ \\[12mm]
$ c_{i,L} $ &  $ \bar{\epsilon}^4\,c_{i,R} $ &  
$\begin{array}{l}  
-\bar{\epsilon}\left(c_{\text{i(3)}}+c_{\text{i(6)}}\right)-\bar{\epsilon}^3\left(c_{\text{i(4)}}+c_{\text{i(5)}}\right) +\\
\\[2mm]
+\bar{\epsilon}^2\left(c_{\text{i(1)}}+c_{\text{i(2)}}+c_{\text{i(7)}}+c_{\text{i(8)}}\right)
\end{array}$  &  $ 24 $ \\[8mm]
\hline \hline
\end{tabular}
\caption{\sf The initial coefficient $c_{i,L}$ from the LH operator $\cP_{i,L}$ (1st column) receives a contribution from the RH $\cP_{i,R}$ (2nd column), plus a combination from the mixing LRH $\cP_{i(j),LR}$ (3rd column), and for each one of the $\cP_{i,L}$ (4th column). The sum of the values at the first, second and third column defines the redefined coefficients $\tilde{c}_{i,L}$ after integrating out $W^\mu_R$ from the physical spectrum. The parameter $\bar{\epsilon}$ stands for $\bar{\epsilon}\equiv \epsilon\,c_{C,LR}$ with $\epsilon\equiv \fL/\fR$.} 
\label{Redefined-coefficients}
}
\end{table}

\nt This will be of relevance for the EWPT parameters $S$ and $T$, as they are sensitive to the effects from $\cP_{1,L}$ and $\cP_{T,L}$ respectively, being thus a testers of the emerging NP effects after removing the RH gauge field content. Furthermore, the triple gauge--boson couplings $\gamma\,W^+\,W^-$ and $W^+\,W^-\,Z$ (TGC) will be also sensitive to the induced effects. In particular, the vertexes $W_\mu^+ W_\nu^- V^{\mu\nu}$, with $V \equiv \{\gamma, Z\}$, will receive non--zero contributions from both of $\{\cP_{1,L},\,\cP_{2,L}\}$ as it will be shown later. Likewise, pair gauge bosons--Higgs couplings will be affected too. In fact, the vertexes $\{F_{\mu\nu} F^{\mu\nu} h,\,Z_{\mu\nu} Z^{\mu\nu} h,\,F_{\mu\nu} Z^{\mu\nu} h,\,Z_\mu\,Z^{\mu\nu}\,\partial_\nu h,\,Z_\mu\, F^{\mu\nu}\,\partial_\nu h\}$, and $\{W_\mu^\dag  W^\mu h,\,Z_\mu Z^\mu h\}$ will depend of linear combinations of the operators in~\eqref{Sensitive-operators}. Additional contributions from the mixing LRH operator $\cP_{C,LR}$ are also found for the interacting terms $\{W_\mu^\dag  W^\mu h,\,Z_\mu Z^\mu h\}$, as it will be described in the next sections.

\nt The contributions to the LH operators in Table~\ref{Redefined-coefficients} are weighted by powers of the parameter $\bar{\epsilon}$ introduced in~\eqref{Gauge-field-EOM}. Two quantities control thus the low energy effects induced by the RH and LRH operators: i) the mixing coefficient $c_{C,LR}$ through the custodial mixing operator $\cP_{C,LR}(h)$; ii) and the scale ratio $\epsilon\equiv \fL/\fR$. In general $\fL \ll \fR$, therefore the small range of $c_{C,LR}$ additionally suppresses the $\epsilon$--effect. Nonetheless, sizeable linear $\bar{\epsilon}$--effects will arise contributing to the LH operators for a $c_{C,LR}$ around its maximal bound together with NP effects not far above the scale $\fL$ as it will be analysed later.

\section{Low energy phenomenology}
\label{pheno}

\nt To analyse the physical impact and the low energy effects sourced by removing the heavy right handed gauge fields, it is necessary to establish the observables that will be set as the parameters input of the model. The renormalization procedure accounting for this is described in the following section.


\subsection{Renormalization scheme}\label{sec:renormalization}

\nt Before integrating out the gauge fields $W^\mu_R$ and with fermion masses neglected from the beginning, the
SM-like Lagrangian $\bar{\LL}_0$ in~\eqref{Lchiral-redefined} (SM-like for the case of $\alpha_B=\alpha_{WB}=\alpha_W=0$)
contains five electroweak parameters $\{g_s,\,g,\,g',\,v,\,\lambda\}$, with the last one as the $h$ self-coupling. The following set of well-measured observables serve to constrain such set of EW parameters and defines the so-called Z-scheme
\begin{itemize}

\item Strong coupling constant $\a_s$, world average~\cite{Beringer:1900zz},
\item Fermi constant $G_F$, extracted from the muon decay rate~\cite{Beringer:1900zz},
\item Fine structure constant $\aem$, extracted from Thomson scattering~\cite{Beringer:1900zz},
\item Gauge boson mass $m_Z$, extracted from the $Z$ lineshape at  LEP I~\cite{Beringer:1900zz},
\item Higg mass $m_h$ now measured at LHC~\cite{Aad:2012tfa,Chatrchyan:2012ufa}.

\end{itemize}

\nt Expressions depending on the parameters $g$, $g'$, $v$ (and $e$) or the weak mixing angle $\theta_W$ will be arranged as combinations of the experimental inputs above. In fact, the electric charge $e$, the weak mixing angle $\theta_W$ and the vev scale $v$ can be defined as 
\be
 e^2 = 4 \pi \a_\text{em}\,, \qquad
 \sin^2\theta_W = \dfrac{1}{2}\left
(1-\left(1-\dfrac{4\pi\a_\text{em}}{\sqrt{2}\GF m_Z^2}\right)^{1/2}\right)\,,\qquad v^2 = \dfrac{1}{\sqrt{2}\GF}\,,
\label{param}
\eeq

\nt and therefore the couplings $g$ and $g'$ as
\be
g = \dfrac{e}{\sin\theta_W}\,,\qquad g'=\dfrac{e}{\cos\theta_W}\,.
\label{param}
\ee

\nt Working in the unitary gauge to analyse the impact that the 
couplings of $\Delta \bar{\LL}_{\text{CP}}$ in~\ref{Lchiral-redefined} have on $\LL_0$, it is
straightforward to show that the set $\{\cP_B,\,\cP_{W,L},\,\cP_G,\,\cP_H,\,\cP_{1,L},\,\cP_{12,L}\}$ introduce corrections to the SM kinetic terms, and in consequence field redefinitions are necessary to obtain canonical kinetic terms. Among the latter operators, $\{\cP_B,\,\cP_{W,L},\,\cP_G\}$ can be considered innocuous operators with respect to $\LL_0$, as the impact on the latter of $c_B$, $c_{W,L}$ and $c_G$ can be totally eliminated from the Lagrangian via field and
coupling constant redefinitions that will impact on certain BSM couplings in $\Delta\LL$ involving external scalar fields. Implementing canonical kinetic terms, it is then easy to identify the
 contribution of $\Delta \bar{\LL}_{\text{CP}}$ to the input parameters\footnote{Following the convention in~\cite{Alonso:2012px,Brivio:2013pma}, BSM
 corrections for the input parameters will be generically denoted by ``$\delta$", whereas the predicted measurable departures from SM expectations will be indicated by ``$\Delta$".}:
\beq
\begin{aligned}
\dfrac{\delta \aem}{\aem}&= 4\,e^2 \left(\tilde{c}_{1,L}+\tilde{c}_{12,L}\right)-s_{2W}\,\alpha_{WB}\,,\\
\dfrac{\delta m_Z}{m_Z}&=  \frac{s_{2W}}{2}\,\alpha_{WB}- \frac{\tilde{c}_{T,L}}{2}-2\,e^2\,\tilde{c}_{1,L}+\frac{2\,e^2 \,c_W^2}{s_W^2}\,\tilde{c}_{12,L}\,.
\end{aligned}
\eeq

\nt where the Fermi constant and the Higgs mass are not corrected by the operators contribution at tree level, and  linear terms in the coefficients $\alpha_{WB}$ and $\tilde{c}_i$ have been kept. All other SM parameter in $\bar{\LL}_\text{chiral}$ can be expressed in terms of the input parameters described above as described in the next.

\subsubsection*{$W$ mass}

\nt Including the effects from the operators in  
$\Delta \bar{\LL}_{\text{CP}}$ in~\eqref{Lchiral-redefined}, the predicted mass departures with respected to the mass value in~\eqref{Charged-masses-expanded} as
\be
\dfrac{\Delta M^2_W}{M^2_W}=
\frac{c_W^2 s_{2W}}{c_{2W}}\,\alpha_{WB}\,\,-\,\,\frac{c_W^2}{c_{2W}} \left[4\,e^2 \left(\tilde{c}_{1,L}-\frac{c_W^2}{s_W^2}\,\tilde{c}_{12,L}\right)+\tilde{c}_{T,L}\right]\,.
\label{W-mass-variation}
\ee

\nt Hereafter the compact notation encoded through the coefficients $c_W$, $s_W$, $c_{2W}$ and $s_{2W}$ will stand for $c_W\equiv \cos\theta_W$, $s_W\equiv \sin\theta_W$, $c_{2W}\equiv \cos\left(2\,\theta_W\right)$ and $s_{2W}\equiv \sin\left(2\,\theta_W\right)$ respectively. Notice two different terms contributing in~\eqref{W-mass-variation}: one accounting for the effects of integrating out the RH fields via $\alpha_{WB}$ in~\eqref{Strength-kinetic-shifts}, and one more accounting for the combined effects from the non-linear operators themselves plus the integration-out of the RH fields, via the redefined operator coefficients $\{\tilde{c}_{T,L},\,\tilde{c}_{1,L},\,\tilde{c}_{12,L}\}$. From Table~\ref{Redefined-coefficients}, the coefficient $\tilde{c}_{T,L}$ receives direct contributions from the set $\{c_{T,L},\,c_{T,R},\,c_{T,LR}\}$, whereas for the hierarchical case $\fL\ll \fR$ (with $g' \ll \gR$), $\tilde{c}_{1,L}$ receives relevant contributions from the LH operator $\cP_{1,L}$ plus a combination from the LRH set $\{\cP_{W,LR},\,\cP_{\text{12(1)}}\}$. The coefficients $\tilde{c}_{12,L}$ only gains contributions from $\cP_{12,L}$. So, all in all the $W$--mass prediction turns out to be sensitive to the operators set
\be
\{\cP_{T,L},\,\cP_{T,R},\,\cP_{1,L},\,\cP_{12,L},\,\cP_{T,LR},\,\cP_{W,LR},\,\cP_{\text{12(1)}}\}\,.
\label{W-mass-Sensitive-operators}
\ee

\nt At the hierarchical limiting case the mass variation in~\eqref{W-mass-variation} becomes
\be
\dfrac{\Delta M^2_W}{M^2_W}=
\frac{c_W^2}{c_{2W}}\left[c_{T,L}-2\,c_{T,LR}+c_{T,R}-e^2\left(4\,c_{1,L}-\frac{4\,c_W^2}{s_W^2}\, c_{12,L}-c_{W,LR}+4\,c_{\text{12(1)}}\right)\right]\,,
\label{W-mass-variation-hierarchical}
\ee

\nt where $\{c_{T,LR},\,c_{W,LR}\}$ are correspondingly the operator coefficient of $\{\cP_{T,LR},\,\cP_{W,LR}\}$ in~\eqref{WT-LR}, whilst $c_{\text{12(1)}}$ is the corresponding one in~\eqref{DeltaL-CP-even-LR} for $\cP_{12(1),LR}$ (see Appendix~\ref{App:Operators-list-LR}).

\subsubsection*{$S$ and $T$ parameters}

\nt Integrating out the right handed gauge fields together with the non--linear operators lead to the tree-level contributions to the oblique parameters $S$ and $T$~\cite{Peskin:1990zt}, as
\be
\aem\,\Delta S =  2\,s_{2W}\,\alpha_{WB}-8\,e^2\,\tilde{c}_{1,L} 
\qquad\qquad \hbox{ and } \qquad\qquad
\aem\,\Delta T =  2\,\tilde{c}_{T,L}\,.
\label{S-T-parameters}
\ee

\nt Notice from~\eqref{S-T-parameters} that only the $S$--parameter is sensitive to the effects of the RH fields integration via $\alpha_{WB}$. Furthermore, combined effects from the non-linear operators plus the removal of $W^\mu_R$ contribute to $S$ via the redefined coefficient $\tilde{c}_{1,L}$. For the hierarchical case $\fL\ll \fR$ (with $g' \ll \gR$), $\tilde{c}_{1,L}$ receives relevant contributions only from the set $\{\cP_{1,\,L},\,\cP_{W,\,LR},\,\cP_{12(1),LR}\}$. In this case the $S$--parameter reduces to
\be
\aem\,\Delta S =  -8\,e^2\,\Big(c_{1,L}\,-\,\frac{1}{4}\,c_{W,LR}\,+\,c_{\text{12(1)}}\Big)\,.
\label{S-parameter-hierarchical}
\ee

\nt The set of custodial breaking operators $\{\cP_{T,\,L},\,\cP_{T,\,R},\,\cP_{T,\,LR}\}$ contribute to $T$ via the coefficient $\tilde{c}_{T,L}$. From Table~\ref{Redefined-coefficients} the $T$--parameters turns to be then
\be
\aem\,\Delta T =  2\,\left(c_{T,L}\,+\,c_{T,R}\,-\,2 c_{T,LR}\right)\,.
\label{T-parameter-hierarchical}
\ee

\nt The experimental values $S= 0.00 \pm 0.10$ and $T=0.02 \pm 0.11$~\cite{Beringer:1900zz} allow to infer the rough order of magnitude estimates $\{c_{1,L},\,c_{W,LR},\,c_{\text{12(1)}}\}\sim 10^{-3}$ and $\{c_{T,L},\,c_{T,R},\,c_{T,LR}\}\sim 10^{-3}$ respectively. More precise ranges for all these coefficients can be derived from a global fit to the EWPD parameters as it will be seen in the next sections.


\subsection{Triple gauge--boson couplings}

\nt The final effective operators contained in $\Delta \bar{\LL}_{\text{CP}}$, weighted by the redefined operators coefficients $\{\tilde{c}_{B},\,\tilde{c}_{i,\chi}\}$, give rise to triple gauge--boson couplings $\gamma\,W^+\,W^-$ and $W^+\,W^-\,Z$ (TGC). These couplings can be generically described through the customary parametrization~\cite{Hagiwara:1986vm} 
\be
\hspace*{-0.3cm}
\begin{aligned}
\frac{\LL_{\text{TGV}}}{g_{WWV}} =& \,i\Bigg\{ 
g_1^V \Big( W^+_{\mu\nu} W^{- \, \mu} V^{\nu} - 
W^+_{\mu} V_{\nu} W^{- \, \mu\nu} \Big) 
   \,+\, \kappa_V W_\mu^+ W_\nu^- V^{\mu\nu}\, +  \frac{ig}{m_W^2} \,\lambda_V\,V^{\mu\nu}\,W^{-\rho}_\mu\, W^+_{\rho\nu} \,+\, \\  \\
&\phantom{- \,i\,\,\,} -  ig_5^V \lambda^{\mu\nu\rho\sigma}
\left(W_\mu^+\partial_\rho W^-_\nu-W_\nu^-\partial_\rho W^+_\mu\right)
V_\s \,+\,
 g_{6}^V \left(\derp_\mu W^{+\mu} W^{-\nu}-\derp_\mu W^{-\mu} W^{+\nu}\right)
V_\nu  \Bigg\}\,,
\label{TGV-Lagrangian}
\end{aligned}
\ee

\nt where $V \equiv \{\gamma, Z\}$ and $g_{WW\gamma} \equiv e$, $g_{WWZ} \equiv e\,c_W/s_W$, and $W^\pm_{\mu\nu}$ and $V_{\mu\nu}$ standing for the kinetic part of the implied gauge field strengths. Electromagnetic gauge invariance requires $g_{1}^{\gamma} =1$ and $g_5^\gamma=0$, and in consequence the CP-even TGC encoded in~\eqref{TGV-Lagrangian} depends in all generality on six
dimensionless couplings $g_1^{Z}$, $g_5^Z$, $g_{6}^{\gamma,Z}$ and
$\kappa_{\gamma,Z}$. Their SM values are $g_1^{Z}=\kappa_{\gamma}=
\kappa_Z=1$ and $g_5^Z=g_{6}^{\gamma}=g_{6}^Z=0$. Couplings
$\lambda_{\gamma,Z}$ turn out to be vanishing up to the $p^4$--order 
for the non--linear treatment assumed in here. As long as CP--even bosonic $p^6$--operators  are considered, a non-vanishing contribution for such couplings is turned on. Additionally, the couplings $g_{6}^V$ have been introduced to account for the contributions associated to the operators containing the
contraction $\DLL_\mu\VLLmuu$, with its corresponding $\partial_\mu\VLLmuu$--part vanishing only for on-shell gauge bosons. 
When fermion masses are neglected, such contraction can be disregarded\footnote{For a general discussion on possible
``off-shell'' vertices associated to $d=4$ and $d=6$ operators see Ref.~\cite{Feruglio:1996bx}.} in the present context (see Appendix~\ref{App:EOM}). 
\begin{table}[htpb!]
\centering
\hspace*{-0.9cm}
\renewcommand{\arraystretch}{1.5}
\begin{tabular}{c||ccc}
\hline\hline
\bf{\text{TGC}} & \bf{\text{SM}} & \bf{\text{Integrating}}  & \bf{\text{Integrating + Operators}} \\
\hline\hline 
$ \mathit{g}_{\text{\textit{$1$}}}^Z $ & $ 1 $ & $ -\frac{2 s_W^4}{c_{2W} s_{2W}}\,\alpha_{WB} $ & $ \frac{1}{2 c_{2W}}\Big(\tilde{c}_{T,L}-4 e^2 \left(\tilde{c}_{12,L}-\frac{s_W^2 \tilde{c}_{1,L}}{c_W^2}\right)\Big)-\frac{4 e^2 \tilde{c}_{3,L}}{s_{2W}^2} $ \\[5mm] 
$ \kappa _{\text{\textit{$\gamma $}}} $ & $ 1 $ & $ \frac{c_W}{s_W}\, \alpha_{WB} $ & $ -\frac{e^2}{s_W^2}\left(2 \tilde{c}_{1,L}+2 \tilde{c}_{2,L}+\tilde{c}_{3,L}+4 \tilde{c}_{12,L}+2 \tilde{c}_{13,L}\right) $ \\[5mm] $
 \kappa _{\text{\textit{$Z$}}} $ & $ 1 $ & $ -\frac{s_{2W}}{2 c_{2W}}\,\alpha_{WB} $ & $ e^2 \left(-\frac{\left(\frac{1}{c_{2W}}+3\right) \tilde{c}_{12,L}+\tilde{c}_{3,L}+2 \tilde{c}_{13,L}}{s_W^2}+\frac{2 \tilde{c}_{1,L}}{c_{2W}}+\frac{2 \tilde{c}_{2,L}}{c_W^2}\right)+\frac{\tilde{c}_{T,L}}{2\,c_{2W}} $ \\[5mm] 
$ \mathit{g}_{\text{\textit{$5$}}}^{\text{\textit{$Z$}}} $ & $ - $ & $ - $ & $ -\frac{4 e^2}{s_{2W}^2}\,\tilde{c}_{14,L} $ \\[5mm] $
 \mathit{g}_{\text{\textit{$6$}}}^{\text{\textit{$\gamma $}}} $ & $ - $ & $ - $ & $ \frac{e^2}{s_W^2}\, \tilde{c}_{9,L} $ \\[5mm] $
 \mathit{g}_{\text{\textit{$6$}}}^Z $ & $ - $ & $ - $ & $ e^2 \left(\frac{4 \tilde{c}_{16,L}}{s_{2 W}^2}-\frac{\tilde{c}_{9,L}}{c_{W}^2}\right) $ \\[5mm]  
\hline \hline
\end{tabular}
\caption{\sf TGV couplings from the Lagrangian $\LL_{\text{TGV}}$ in~\eqref{TGV-Lagrangian} following standard convention in Ref.~\cite{Hagiwara:1986vm}. The total TGV coupling (1st column)  is made out of: the usual SM  contribution (2nd column) + additional effects after integrating out the fields $W^\mu_R$ (3rd column) + the combined terms yielded by the RH fields removal and the non-linear operators all together (4th column).} 
\label{TGV-couplings-BSM}
\end{table}

\nt The set of TGC parametrized through $\LL_{\text{TGV}}$ in~\eqref{TGV-Lagrangian} are listed in Table~\ref{TGV-couplings-BSM} (1st column), being split all of them into their corresponding SM  contribution (2nd column), the additional effect after integrating out the gauge fields $W^\mu_R$ (3rd column), plus the combined terms accounting for the combined effect by removing the RH fields and the non-linear operators via redefined coefficients (4th column). From Tables~\ref{Redefined-coefficients}-\ref{TGV-couplings-BSM} and the operators set in~\eqref{Sensitive-operators}-\eqref{Left-Right-operators}, it is inferred that

\begin{itemize}

\item The TGC set $\{\mathit{g}_{\text{\textit{$1$}}}^Z ,\,\kappa _{\text{\textit{$\gamma $}}},\,\kappa _{\text{\textit{$Z$}}}\}$ depends, among others, on $\{\tilde{c}_{1,L},\,\tilde{c}_{2,L}\}$, being sensitive therefore to the contributions from $\{\cP_{W,LR},\,\cP_{\text{12(1)}},\,\cP_{\text{3(2)}},\,\cP_{\text{13(2)}}\}$ for the hierarchical case as it was mentioned before.

\item No effects are induced onto the TGC set $\{g_5^Z,\,g_{6}^{\gamma},\,g_{6}^Z\}$ after integrating out the RH fields, only the combined effects from the removal plus the non-linear operators through the redefined coefficients. No RH operators neither LRH mixing ones contribute to them.

\end{itemize}

\subsection{Quartic gauge--boson couplings}

\nt The quartic gauge-boson couplings (QGC) can be parametrized in the following Lagrangian
\be
\begin{aligned}
\LL_{\text{QGV}}& \,=\, g^2\Big\{g^{(1)}_{WWWW}\,W_{\mu}^{\dagger}\,W^{\mu\dagger} \,W^{\nu}\,W_{\nu}\,\,-\,\,g^{(2)}_{WWWW}\,\Big(W_{\mu}^{\dagger} \,W^{\mu}\Big)^2\,\,+\,\,g_{ZZZZ}\,\Big(Z^\mu\,Z_\mu\Big)^2\,\,+\,\,\\ 
&\phantom{g^2\,\,\,\,}\,\,-\,\, g^{(1)}_{VVWW}\,V^\mu\,V_\mu\,W^\dag_\nu\,W^\nu\,\,+\,\,g^{(2)}_{VVWW}\,V^{\mu}\,V_{\nu}\,W_{\mu}^{\dagger}\,W^{\nu} \,\,-\,\, g^{(1)}_{\gamma WWZ}\,A^\mu\,Z_\mu\,W^\dag_\nu\,W^\nu\,\,+
\\ 
&\phantom{g^2\,\,\,\,}\,\,+\,\,\left(g^{(2)}_{\gamma WWZ}\,A^{\mu}\,Z_{\nu}\, W_{\mu}^{\dagger}\,W^{\nu}\,\,+\,\,\hc\right)\,\,+\,\,i\,g^{(3)}_{\gamma WWZ}\,\e^{\mu\nu\rho\s}\,W^+_\mu\,W^-_\nu\,A_\rho \,Z_\s\Big\}
\label{Quartic-gauge-interactions}
\end{aligned}
\ee

\nt where again $V \equiv \{\gamma, Z\}$. All these couplings are gathered up in Table~\ref{Quartic-gauge-couplings}, where a similar splitting for each one of them is done alike to the previous TGC case. 
\begin{table}[htb!] 
\small{
\centering
\hspace*{-0.5cm}
\renewcommand{\arraystretch}{0.0}
\begin{tabular}{c||ccc}
\hline\hline
\bf{\text{QGC}} & \bf{\text{SM}}  & \bf{\text{Integrating}}  & \bf{\text{Integrating + Operators}} \\
\hline\hline 
\\[1mm]
$ \mathit{g}_{\text{\textit{$W W W W$}}}^{(1)} $ & $ \frac{1}{2} $ & $ -\frac{c_W s_W^3}{c_{2W}}\,\alpha_{WB} $ & $ \frac{e^2 \left(4 s_W^4 \tilde{c}_{1,L}+c_{2W} \left(\left(c_{2W}-8\right) \tilde{c}_{12,L}-2 \tilde{c}_{3,L}+\tilde{c}_{11,L}-4 \tilde{c}_{13,L}\right)-\tilde{c}_{12,L}\right)}{2 c_{2W} s_W^2}+\frac{s_{2W}^2 \tilde{c}_{T,L}}{8 c_{2W} s_W^2} $ \\[5mm] $
 \mathit{g}_{\text{\textit{$W W W W$}}}^{(2)} $ & $ \frac{1}{2} $ & $ -\frac{c_W s_W^3}{c_{2W}}\, \alpha_{WB} $ & $ \frac{32 e^2 s_W^4 \tilde{c}_{1,L}+\left(c_{4w}-1\right) \left(4 e^2 \tilde{c}_{12,L}-\tilde{c}_{T,L}\right)-8 e^2 c_{2W} \left(2 \tilde{c}_{3,L}+2 \tilde{c}_{6,L}+\tilde{c}_{11,L}+8 \tilde{c}_{12,L}+4 \tilde{c}_{13,L}\right)}{16 c_{2W} s_W^2} $ \\[5mm] $
 \mathit{g}_{\text{\textit{$Z Z Z Z$}}} $ & $ - $ & $ - $ & $ \frac{e^2}{4 c_W^4 s_W^2}\left(\tilde{c}_{6,L}+\tilde{c}_{11,L}+2 \left(\tilde{c}_{23,L}+\tilde{c}_{24,L}+2 \tilde{c}_{26,L}\right)\right) $ \\[5mm] 
$ \mathit{g}_{\text{\textit{$\gamma  \gamma  W W$}}}^{(1)} $ & $ s_W^2 $ & $ -\frac{s_{2W}^3}{4 c_{2W}}\,\alpha_{WB} $ & $ \frac{1}{4 c_{2W}}\Big(s_{2W}^2 \left(4 e^2 \tilde{c}_{1,L}+\tilde{c}_{T,L}\right)-16 e^2 c_W^4 \tilde{c}_{12,L}\Big) $ \\[5mm] 
$ \mathit{g}_{\text{\textit{$W W Z Z$}}}^{(1)} $ & $ c_W^2 $ & $ -\frac{2 c_W s_W^5}{c_{2W}}\,\alpha_{WB} $ & $ \frac{4 e^2 s_W^4 \tilde{c}_{1,L}}{c_{2W}}-e^2 \left(\frac{s_{2W}^2 \tilde{c}_{12,L}}{c_{2W}}+\frac{2 \tilde{c}_{3,L}}{s_W^2}+\frac{4 \left(\tilde{c}_{6,L}+\tilde{c}_{23,L}\right)}{s_{2W}^2}\right)+\frac{c_W^4 \tilde{c}_{T,L}}{c_{2W}} $ \\[5mm] 
$ \mathit{g}_{\text{\textit{$\gamma  \gamma  W W$}}}^{(2)} $ & $ s_W^2 $ & $ -\frac{s_{2W}^3}{4 c_{2W}}\,\alpha_{WB} $ & $ \frac{1}{4 c_{2W}}\Big(s_{2W}^2 \left(4 e^2 \tilde{c}_{1,L}+\tilde{c}_{T,L}\right)-4 e^2 \left(4 c_W^4 \tilde{c}_{12,L}+c_{2W} \tilde{c}_{9,L}\right)\Big) $ \\[8mm] 
$ \mathit{g}_{\text{\textit{$W W Z Z$}}}^{(2)} $ & $ c_W^2 $ & $ -\frac{2 c_W s_W^5}{c_{2W}}\,\alpha_{WB} $ & 
$\begin{array}{l}  
\frac{1}{c_{2W}}\Big(\frac{4 e^2 \left(s_W^6 \tilde{c}_{1,L}-c_W^6 \tilde{c}_{12,L}\right)}{s_W^2}+c_W^4 \tilde{c}_{T,L}\Big)+\\
\\[1mm]
+e^2 \left(-\frac{s_W^2 \tilde{c}_{9,L}}{c_W^2}+\frac{4 \left(\tilde{c}_{11,L}+\tilde{c}_{24,L}\right)}{s_{2W}^2}+\frac{4 c_W^2 \tilde{c}_{12,L}-2 \tilde{c}_{3,L}}{s_W^2}+\frac{2 \tilde{c}_{16,L}}{c_W^2}\right) 
\end{array}$
\\[12mm] 
$ \mathit{g}_{\text{\textit{$\gamma  W W Z$}}}^{(1)} $ & $ s_{2W} $ & $ -\frac{\left(c_{4w}+3\right) s_W^2}{2 c_{2W}}\,\alpha_{WB} $ & $ \frac{1}{c_{2W}}\Big(e^2 \left(c_{4w}+3\right) \left(\frac{s_W \tilde{c}_{1,L}}{c_W}-\frac{c_W \tilde{c}_{12,L}}{s_W}\right)+2 c_W^3 s_W \tilde{c}_{T,L}\Big)-\frac{4 e^2 \tilde{c}_{3,L}}{s_{2W}} $ \\[8mm] 
$ \mathit{g}_{\text{\textit{$\gamma  W W Z$}}}^{(2)} $ & $ \frac{1}{2}s_{2W} $ & $ -\frac{\left(c_{4w}+3\right) s_W^2}{4 c_{2W}}\,\alpha_{WB} $ & 
$\begin{array}{l}  
\frac{1}{2 c_{2W}}\Big(e^2 \left(-\frac{4 c_{2W} \tilde{c}_{3,L}}{s_{2W}}-\frac{c_W \left(\left(c_{4w}+3\right) \tilde{c}_{12,L}+2 \tilde{c}_{16,L}\right)}{s_W}+\right.\\
\\[-1mm]
\left.+\frac{s_W \left(\left(c_{4w}+3\right) \tilde{c}_{1,L}+2 \left(c_{2W} \tilde{c}_{9,L}+\tilde{c}_{16,L}\right)\right)}{c_W}\right)+2 c_W^3 s_W \tilde{c}_{T,L}\Big) 
\end{array}$
\\[8mm] 
$ \mathit{g}_{\text{\textit{$\gamma  W W Z$}}}^{(3)} $ & $ - $ & $ - $ & $ -\frac{2 e^2 \tilde{c}_{14,L}}{s_{2W}} $ \\[3mm] 
\hline \hline
\end{tabular}
\caption{\sf QGC values from $\LL_{\text{QGV}}$ in~\eqref{Quartic-gauge-interactions}. All the couplings are split as in Table~\ref{TGV-couplings-BSM}.} 
\label{Quartic-gauge-couplings}
}
\end{table}
Likewise it is inferred:
\begin{itemize}

\item All the QGC, but $\{\mathit{g}_{\text{\textit{$Z Z Z Z$}}},\,\mathit{g}_{\text{\textit{$\gamma  W W Z$}}}^{(3)}\}$, have a dependence on $\{\tilde{c}_{T,L},\,\tilde{c}_{1,L}\}$, and thus on the contributions from $\{\cP_{T,R},\,\cP_{T,LR},\,\cP_{W,LR},\,\cP_{\text{12(1)}}\}$ for the hierarchical case.

\item The set $\{\mathit{g}_{\text{\textit{$Z Z Z Z$}}},\,\mathit{g}_{\text{\textit{$\gamma  W W Z$}}}^{(3)}\}$ depends only on the combined effects through the corresponding redefined coefficients in Table~\ref{Quartic-gauge-couplings}, being sensitive to the LH operators only.

\end{itemize}


\subsection{Triple gauge--$h$ couplings}

\nt $\Delta \bar{\LL}_{\text{CP}}$ gives rise to the cubic gauge-Higgs interactions, here encoded in the Lagrangian as
\be
\begin{aligned}
&\LL_{hVV} = \frac{1}{v}\Big\{g_{\gamma\gamma h} F_{\mu\nu} F^{\mu\nu} h
+ g_{ hZZ}^{(1)} Z_{\mu\nu} Z^{\mu\nu} h
+ \mathit{g}^{(1)}_{\gamma  h Z} F_{\mu\nu} Z^{\mu\nu} h
+ g_{hWW}^{(1)} W^\dag_{\mu\nu}W^{\mu\nu}h +
\\ 
& + g_{hZZ}^{(2)} Z_\mu\,Z^{\mu\nu}\,\partial_\nu h
+ g^{(3)}_{hZZ}\,\partial^\mu Z_\mu\,Z^{\nu}\,\partial_\nu h + \mathit{g}_{\gamma h Z}^{(2)} Z_\mu\, F^{\mu\nu}\,\partial_\nu h
+ \left(g^{(2)}_{hWW} W^\mu\,W^\dag_{\mu\nu}\,\partial^\nu h + {\rm h.c.} \right)  +
\\
& + \left(g^{(3)}_{hWW}\,\partial^\mu W^\dag_{\mu}\,W^\nu\,\partial_\nu h + {\rm h.c.} \right)  + g^{(4)}_{hZZ}\,\partial^\mu Z_\mu\,\partial^\nu Z_{\nu}\, h + g^{(4)}_{hWW}\,\partial^\mu W^\dag_{\mu}\,\partial_\nu W^\nu\, h  +
\\
&
+ g_{hWW}^{(5)}\,W_\mu^\dag  W^\mu h 
+ g_{hWW}^{(6)}\,W_\mu^\dag\,W^\mu\,\Box h
+ g_{hZZ}^{(5)}\,Z_\mu Z^\mu h
+ g_{hZZ}^{(6)}\,Z_\mu Z^\mu\,\Box h\Big\}\,.
\label{Cubic-gauge-h-interactions}
\end{aligned}
\ee

\nt Table~\ref{Cubic-gauge-h-couplings} collects all the couplings. All them have no contributions after integrating out the RH fields, but $\{\mathit{g}_{\text{\textit{$h W W$}}}^{(5)},\,\mathit{g}_{\text{\textit{$h Z Z$}}}^{(5)}\}$, whereas the set $\{\mathit{g}_{\text{\textit{$\gamma  \gamma  h$}}}^{(1)},\,\mathit{g}_{\text{\textit{$h Z Z$}}}^{(1)},\,\mathit{g}_{\text{\textit{$\gamma  h Z$}}}^{(1)},\,\mathit{g}_{\text{\textit{$h W W$}}}^{(5)},\,\mathit{g}_{\text{\textit{$h Z Z$}}}^{(5)}\}$ is sensitive to the operators $\{\cP_{C,R},\,\cP_{W,R},\,\cP_{1,R},\,\cP_{12,R},\,\cP_{T,R}\}$ and $\{\cP_{C,LR},\,\cP_{T,LR},\,\cP_{W,LR},\,\cP_{\text{12(1)}}\}$ for the hierarchical case, according to the involved coefficient dependence in Table~\ref{Redefined-coefficients}.\\

\begin{table}[htb!]
\small{
\centering
\renewcommand{\arraystretch}{0.0}
\begin{tabular}{c||ccc}
\hline\hline
\bf{\text{hVV}} & \bf{\text{SM}} & \bf{\text{Integrating}}  & \bf{\text{Integrating + Operators}} \\
\hline\hline 
$ \mathit{g}_{\text{\textit{$\gamma  \gamma  h$}}}^{(1)} $ & $ - $ & $ - $ & $ -\frac{1}{2} e^2 \left(\tilde{a}_{W,L}-4 \left(\tilde{a}_{1,L}+\tilde{a}_{12,L}\right)+\tilde{a}_B\right) $ \\[3mm] $
 \mathit{g}_{\text{\textit{$h Z Z$}}}^{(1)} $ & $ - $ & $ - $ & $ -\frac{1}{2} e^2 \left(\frac{c_W^2 \left(\tilde{a}_{W,L}-4 \tilde{a}_{12,L}\right)}{s_W^2}+4 \tilde{a}_{1,L}+\frac{\tilde{a}_B s_W^2}{c_W^2}\right) $ \\[3mm] $
 \mathit{g}_{\text{\textit{$\gamma  h Z$}}}^{(1)} $ & $ - $ & $ - $ & $ e^2 \left(\frac{c_W \left(4 \tilde{a}_{12,L}-\tilde{a}_{W,L}\right)}{s_W}+\frac{4 c_{2W} \tilde{a}_{1,L}}{s_{2W}}+\frac{\tilde{a}_B s_W}{c_W}\right) $ \\[1mm] $
 \mathit{g}_{\text{\textit{$h W W$}}}^{(1)} $ & $ - $ & $ - $ & $ -\frac{e^2}{s_W^2}\, \tilde{a}_{W,L} $ \\[2mm] $
 \mathit{g}_{\text{\textit{$h Z Z$}}}^{(2)} $ & $ - $ & $ - $ & $ e^2 \left(\frac{2 \tilde{a}_{4,L}}{c_W^2}-\frac{\tilde{a}_{5,L}+2 \tilde{a}_{17,L}}{s_W^2}\right) $ \\[4mm] 
$ \mathit{g}_{\text{\textit{$\gamma  h Z$}}}^{(2)} $ & $ - $ & $ - $ & $ -\frac{2 e^2}{s_{2W}}\left(2 \tilde{a}_{4,L}+\tilde{a}_{5,L}+2 \tilde{a}_{17,L}\right) $ \\[3mm] 
$ \mathit{g}_{\text{\textit{$h W W$}}}^{(2)} $ & $ - $ & $ - $ & $ -\frac{e^2}{s_W^2}\,\tilde{a}_{5,L} $ \\[3mm] 
$ \mathit{g}_{\text{\textit{$h Z Z$}}}^{(3)} $ & $ - $ & $ - $ & $ -\frac{4 e^2}{s_{2W}^2}\left(\tilde{a}_{10,L}+2 \tilde{a}_{19,L}\right) $ \\[4mm] 
$ \mathit{g}_{\text{\textit{$h Z Z$}}}^{(4)} $ & $ - $ & $ - $ & $ -\frac{4 e^2}{s_{2W}^2}\left(\tilde{a}_{9,L}+2 \tilde{a}_{15,L}\right) $ \\[4mm] $
 \mathit{g}_{\text{\textit{$h W W$}}}^{(3)} $ & $ - $ & $ - $ & $ -\frac{e^2}{s_W^2}\,\tilde{a}_{10,L} $ \\[3mm] 
$ \mathit{g}_{\text{\textit{$h W W$}}}^{(4)} $ & $ - $ & $ - $ & $ -\frac{2\,e^2}{s_W^2}\,\tilde{a}_{9,L} $ \\[3mm] 
$ \mathit{g}_{\text{\textit{$h W W$}}}^{(5)} $ & $ 2 c_{W}^2 $ & $ -2 c_{W}^2 s_{2 W} \alpha_{WB} $ & $ c_{W}^2 \left(\left(\tilde{a}_{C,L}-\tilde{c}_{C,L}\right) -\frac{4 e^2 c_{W} s_{2 W} \tilde{c}_{12,L}}{s_{W}^3}+8 e^2 \tilde{c}_{1,L}+2 \tilde{c}_{T,L}-c_H\right) $ \\[4mm] 
$ \mathit{g}_{\text{\textit{$h W W$}}}^{(6)} $ & $ - $ & $ - $ & $ -\frac{2 e^2}{s_W^2}\, \tilde{a}_{7,L} $ \\[4mm] 
$ \mathit{g}_{\text{\textit{$h Z Z$}}}^{(5)} $ & $ 1 $ & 
  $ s_{2 W} \alpha_{WB} $ & 
$ \frac{1}{2} \left(-2 \bar{c}_{C,\text{LR}}+\tilde{a}_{C,L} -2 \tilde{a}_{T,L} + 8 e^2 \left(\frac{c_{W}^2 \tilde{c}_{12,L}}{s_{W}^2}-\tilde{c}_{1,L}\right)-c_H\right) $ 
 \\[4mm] 
$ \mathit{g}_{\text{\textit{$h Z Z$}}}^{(6)} $ & $ - $ & $ - $ & $ -\frac{4 e^2}{s_{2W}^2}\,\left(\tilde{a}_{7,L}+2 \tilde{a}_{25,L}\right) $ \\[4mm]   
\hline \hline
\end{tabular}
\caption{\sf Triple gauge-Higgs couplings from $\LL_{hVV}$ in~\eqref{Cubic-gauge-h-interactions}. The notation $\tilde{a}_i$ stands for $\tilde{a}_i\equiv \tilde{c}_i\,a_i $, with $\tilde{c}_i$ the redefined operator coefficients and $a_i$ from the $\cF(h)$--definition of~\eqref{F}.} 
\label{Cubic-gauge-h-couplings}
}
\end{table}

\boldmath
\subsection{Current operators bounds}
\label{numerical}
\unboldmath

\subsubsection{$S$ and $T$ parameters bounds}

\nt The parameters $S$, $T$, $U$ are precisely obtained through a global fit to electroweak precision data, resulting in the current values and correlation matrix~\cite{Beringer:1900zz}
\begin{eqnarray}
\Delta S=0.00\pm 0.10\,, & \Delta T=0.02\pm 0.11\,, & \Delta U=0.03\pm 0.09\,,\\
& & \nonumber \\
&\rho=\left(\begin{array}{ccc}
1 & 0.89 & -0.55 \\
0.89 & 1 &-0.8 \\
-0.55 & -0.8 &1 
\end{array}\right)\,.\\ \nn
\label{eq:STUexp} 
\end{eqnarray}

\nt Tree-level contributions from the sets $\{\cP_{1,\,L},\,\cP_{W,\,LR},\,\cP_{12(1),LR}\}$ and $\{\cP_{T,\,L},\,\cP_{T,\,R},\,\cP_{T,\,LR}\}$ are generated for the $S$ and $T$ parameters correspondingly (see Eqs.~\eqref{S-parameter-hierarchical}-\eqref{T-parameter-hierarchical}). The 95\% CL allowed ranges for their corresponding coefficients were found in~\cite{Brivio:2013pma}, and can be translated in here, according to~\eqref{S-T-parameters}, as
\begin{equation}
  -4.7\times 10^{-3} \leqslant \tilde{c}_{1,L}-\frac{s_{2W}}{4\,e^2}\,\alpha_{WB}\leqslant 4\times 10^{-3} 
\;\;\; \hbox{ and } \;\;\;
-2\times 10^{-3} \leqslant \tilde{c}_{T,L} \leqslant 1.7 \times 10^{-3}  \;.
\label{S-T-bounds-redefined}
\end{equation}

\nt Through the redefined coefficients in Table~\ref{Redefined-coefficients}, allowed ranges for each one of the involved coefficients in~\eqref{S-parameter-hierarchical}-\eqref{T-parameter-hierarchical} are obtained from~\eqref{S-T-bounds-redefined} and are shown in Table~\ref{S-T-bounds}. 
%
%
\begin{table}[htb!]
\centering
\small{
\renewcommand{\arraystretch}{1.0}
\begin{tabular}{c||c}
\hline\hline
\bf Coefficients & \bf Ranges (95\% CL)  \\
\hline\hline 
$\alpha_{WB}$ &  $[-0.0017,\,0.002]$\\[1mm]  
\hline
$c_{1,L}$ &  $[-0.0047,\,0.004]$\\[2mm]  
$c_{T,L}$ &  $[-0.002,\,0.0017]$\\[2mm]  
\hline
$c_{T,R}$ &  $[-0.002,\,0.0017]$\\[1mm]  
\hline
$c_{T,LR}$ &  $[-0.0008,\,0.001]$\\[2mm]  
$c_{W,LR}$ &  $[-0.016,\,0.018]$\\[2mm]  
$c_{\text{12(1)}}$ &  $[-0.0047,\,0.004]$\\[1mm]  
\hline \hline
\end{tabular}
\caption{\sf Bounds at 95\% CL for each one of the operator coefficients in~\eqref{S-parameter-hierarchical}-\eqref{T-parameter-hierarchical} obtained from~\eqref{S-T-bounds-redefined} (via Table~\ref{Redefined-coefficients}). The bound for $\alpha_{WB}$ and for each one of the coefficients involved in the definition of $\tilde{c}_{1,L}$, i.e $\{c_{1,L},\,c_{W,LR},\,c_{\text{12(1)}}\}$, are obtained by setting to zero the rest of them in~\eqref{S-T-bounds-redefined} (1st inequality). The same comment applies for $\{c_{T,L},\,c_{T,R},\,c_{T,LR}\}$ (2nd inequality).}  
\label{S-T-bounds}
}
\end{table}

\nt The constrains in Table~\ref{S-T-bounds} signal small contributions of the operators $\{\cP_{1,\,L},\,\cP_{W,\,LR},\,\cP_{12(1),LR}\}$ and $\{\cP_{T,\,L},\,\cP_{T,\,R},\,\cP_{T,\,LR}\}$ to the gauge-boson self-couplings and to the present Higgs data. Consequently they will not be included in the following discussion.

\subsubsection{TGC bounds}

\nt Bounds for the TGC can be determined from the two--dimensional analysis in Ref.~\cite{LEPEWWG}, which was performed in terms of the induced variations of the couplings $\kappa_\gamma$, $g_1^Z$ and $\kappa_Z$. Such variations are corresponding in this scenario to the sum of the values at the 3rd and 4th columns in Table~\ref{TGV-couplings-BSM}, here denoted as $\Delta\kappa_\gamma$, $\Delta g_1^Z$ and $\Delta\kappa_Z$ respectively, and satisfying the relations in Appendix~\ref{App:De-correlation-formulae}. From the two--dimensional	 analysis in Ref.~\cite{LEPEWWG} it was obtained
\begin{equation}
\kappa_\gamma=0.984^{+0.049}_{-0.049} \;\;\; \hbox{ and } \;\;\; 
g_1^Z=1.004^{+0.024}_{-0.025}\,,
\label{eq:tgvdata}
\end{equation}

\nt with a correlation factor of $\rho=0.11$. The corresponding 90\%
CL ranges for the coefficients $c_{2,L}$ and $c_{3,L}$ from the TGC data were also found, traded here by $\tilde{c}_{2,L}$ and $\tilde{c}_{3,L}$, and finally translated to the coefficients shown in Table~\ref{TGC-bounds} (via Table~\ref{Redefined-coefficients}).
\begin{table}[htb!]
\centering
\small{
\renewcommand{\arraystretch}{1.0}
\begin{tabular}{c||c}
\hline\hline
\\[-4mm]
\bf Coefficients & \bf Ranges (90\% CL)
\\[0.5mm]
\hline\hline 
\\[-4mm]
$c_{2,L}$ &  $[-0.12,\,0.076]$\\[2mm]  
$c_{3,L}$ &  $[-0.064,\,0.079]$\\[1mm]  
\hline
\\[-4mm]
$c_{\text{3(2)}}$ &  $[-0.24,\,0.152]$\\[2mm]  
$c_{\text{13(2)}}$ &  $[-0.12,\,0.076]$\\[1mm]  
\hline \hline
\end{tabular}
\caption{\sf Bounds at 90\% CL for each one of the operator coefficients $\{\tilde{c}_{2,L},\,\tilde{c}_{3,L}\}$, being translated into bounds for $\{c_{2,L},\,c_{3,L}\}$ and $\{c_{\text{3(2)}},\,c_{\text{13(2)}}\}$ according to Table~\ref{Redefined-coefficients}. } 
\label{TGC-bounds}
}
\end{table}
\nt Finally, notice from Table~\ref{TGV-couplings-BSM} that $g_5^Z$ is generated by the operator $\cP_{14,L}$ only. The best current bounds on this anomalous TGC come from $W^+W^-$ pairs studies and $W$ production at LEP II energies~\cite{Abbiendi:2003mk,Achard:2004ji,Schael:2004tq}. Furthermore, the strongest limits on $g_5^Z$ originate from its impact on radiative corrections to  $Z$ physics~\cite{Eboli:1994jh,Dawson:1994qh,Eboli:1998hb}. In Ref.~\cite{Brivio:2013pma} were reported the available direct and indirect limits on $g_5^Z$. The $90\%$ CL region  from indirect bounds~\cite{Eboli:1994jh,Dawson:1994qh,Eboli:1998hb} turns out to be $g_5^Z\,\,\in\,\,[-0.08,0.04]$, and translated into the corresponding bound for $\cP_{14,L}$ as\footnote{These limits were obtained
assuming only a non-vanishing $g_5^Z$ while the rest of anomalous TGV
were set to their corresponding SM value~\cite{Brivio:2013pma}.}
\be
c_{14,L}\,\,\in\,\,[-0.04,0.02]\,.
\label{c14}
\ee
\nt No extra terms contribute in the hierarchical limiting case to the
coefficient $c_{14,L}$. However, linear $\bar{\epsilon}$--effects would arise from NP not far above the EW scale ($ \epsilon\sim 1$) and for a LRH mixing strength from $\cP_{C,LR}$ close to its maximal bound in~\eqref{cLR-bound}. These contributions and their constrains will be analysed later.
\subsubsection{Bounds on $hVV$ couplings}

\nt In Ref.~\cite{Brivio:2013pma}, the operators set $\{\cP_G,\,\cP_B,\,\cP_{W,L},\,\cP_{C,L},\,\cP_H,\,\cP_{4,L},\,\cP_{5,L}\}$ were bounded from the constraints of the Higgs data on $hVV$--couplings. Notice from Table~\ref{Cubic-gauge-h-couplings} that all them contribute to some of the listed $hVV$--couplings. The set $\{\cP_{7,L},\,\cP_{9,L},\,\cP_{10,L}\}$ was not included in the parameter fit as they do not entail a physical impact on the observables considered, whereas the operators set $\{\cP_{12,L},\,\cP_{17,L},\,\cP_{19,L},\,\cP_{25,L}\}$ does not contribute relevantly for a non-linear realization of the underlying dynamics (see~\cite{Brivio:2013pma} for further details). 

\nt The analysis relies in two operator subsets: in the first one, $\cP_{C,L}$ is neglected, whilst for  the second one its contribution is connected to that of $\cP_H$. Furthermore, the sensitivity of the results to the sign of the $h$-fermion couplings, e.g. $s_Y h\left(\bar{Q}_L\UH \cY_Q Q_R+\hc\right)$, is also considered by performing the analysis with both values  of the discrete parameter $s_Y=\pm$. The detailed discussion on the performed six-parameter fit, and the implied chi--square analysis using the available data on the signal strengths $\mu$, by accounting for the data from Tevatron D0 and CDF Collaborations, as well as from the LHC, CMS, and ATLAS Collaborations at 7 TeV and 8 TeV for final states $\gamma\gamma$, $W^+W^-$, $ZZ$, $Z\gamma$, $b\bar b$, and
$\tau\bar\tau$ \cite{Tuchming:2013wja,atlastau,atlasb,atlaszz,atlasww,
  atlasgamgam,atlasgamgamnew,CMStau,CMSb,CMSb2,CMSzz,CMSww,CMSgamgam,
  Chatrchyan:2013vaa} is referred to the reader in~\cite{Brivio:2013pma} and~\cite{Corbett:2012dm,Corbett:2013pja} for details on the Higgs data analysis.
\begin{table}[htb!]
\centering
\small{
\renewcommand{\arraystretch}{1.0}
\begin{tabular}{c||c||c}
\hline\hline
\\[-4mm]
\bf Coefficients & \bf Ranges (90\% CL) & \bf Ranges (95\% CL)
\\[0.5mm]
\hline\hline 
$c_H$ &  $[-0.66, 0.66]$,\,\,$[-1.1, 0.49]$ &  $[-0.5, 0.6]$\\[2mm]  
$c_B$ &  $[-0.50, 0.21]$ &  $[-0.30, 0.25]$\\[2mm]  
\hline
$c_{W,L}$ &  $[-0.12, 0.51]$ &  $[-0.12, 0.37]$\\[2mm]  
$c_{4,L}$ &  $[-0.47, 0.14]$ &  $[-0.35, 0.12]$\\[2mm]  
$c_{5,L}$ &  $[-0.33, 0.17]$ &  $[-0.13, 0.12]$\\[2mm]  
\hline
$c_{W,R}$ &  $[-0.50, 0.21]$ &  $[-0.30, 0.25]$\\[2mm]  
$c_{1,R}$ &  $[-0.053, 0.13]$ &  $[-0.062, 0.075]$\\[2mm]  
$c_{12,R}$ &  $[-0.053, 0.13]$ &  $[-0.062, 0.075]$\\[2mm]  
\hline
$c_{\text{17(2)}}$ &  $[-0.47, 0.14]$ &  $[-0.35, 0.12]$\\[1mm]  
\hline \hline
\end{tabular}
\caption{\sf 90\% and 95\% CL allowed ranges (2nd and 3rd columns) for the coefficients $\{c_H,\,c_B,\,c_{W,L},\,c_{4,L},\,c_{5,L}\}$, $\{c_{W,R},\,c_{1,R},\,c_{12,R}\}$ and $c_{\text{17(2)}}$ contributing to Higgs data. The 90\% CL ranges for $c_H$ correspond to the two analysis in~\cite{Brivio:2013pma} and for both signs of $s_Y$, whereas the reported ranges for $\{c_{W,L},\,c_{4,L},\,c_{5,L}\}$ are basically the same for both of the operators subset and for both values of $s_Y$. The 95\% CL ranges correspond to the recent SFitter analysis in~\cite{Corbett:2015mqf}.}
\label{hVV-bounds}
}
\end{table}

\nt The 90$\%$ CL allowed ranges for the coefficients $\{c_H,\,\tilde{a}_B,\,\tilde{a}_{W,L},\,\tilde{a}_{4,L},\,\tilde{a}_{5,L}\}$\footnote{The corresponding bound for the gluonic operator coefficient $\a_G$, not listed here neither written down in the gauge-higgs Lagrangian of~\eqref{Cubic-gauge-h-interactions}, can be read off from~\cite{Brivio:2013pma}. In addition it has been assumed $a_i\approx\cO(1)$ in the definition $\tilde{a}_i\equiv \tilde{c}_i\,a_i $.} can be straightforwardly obtained from the corresponding ones in~\cite{Brivio:2013pma}. The latter ranges can be translated (via Table~\ref{Redefined-coefficients}) into ranges for the coefficients $\{c_H,\,c_B,\,c_{W,L},\,c_{4,L},\,c_{5,L}\}$, $\{c_{W,R},\,c_{1,R},\,c_{12,R}\}$ and $c_{\text{17(2)}}$, shown in Table~\ref{hVV-bounds}. The recent SFitter analysis for a non-linear framework in~\cite{Corbett:2015mqf} allows us also to obtain the corresponding 95$\%$ CL allowed ranges collected in Table~\ref{hVV-bounds}. As suggested in~\cite{Brivio:2013pma}, the sensitivity to the coefficients $\{c_H,\,\tilde{a}_B,\,\tilde{a}_{W,L},\,\tilde{a}_{4,L},\,\tilde{a}_{5,L}\}$ can improve by a factor $\cO(3-5)$ with a similar analysis, according to the expected uncertainties accesible in the Higgs signal strengths from ATLAS and CMS at 14 TeV, and for an integrated luminosity of 300 fb$^{-1}$~\cite{CMS:2013xfa,ATLAS:2013hta}. Consequently, a similar  sensitivity improvement is expected for all the coefficients in the Table~\ref{hVV-bounds}, as they are linked to the redefined coefficients through the relations in Table~\ref{Redefined-coefficients}. Finally, the 95$\%$ CL ranges from the SFitter analysis are a bit constraining with respect to the 90$\%$ CL ranges in Table~\ref{hVV-bounds}. 

\nt In summary, the current bounds from the EWPD analysis for the $S$ and $T$ parameters, as well as the TGC and $hVV$--couplings bounds, have allowed us to constraint at the thousandth level the coefficients in Table~\ref{S-T-bounds} (few percent level for $c_{W,LR}$), at the 10\% level those in Table~\ref{TGC-bounds} (percent level for $c_{3,L}$), and at 10\% (percent level for $\{c_{1,R},\,c_{12,R}\}$) the coefficients in Table~\ref{hVV-bounds}. Finally, through the aforementioned experimental current bounds we have obtained bounds for the coefficients weighting correspondingly some of the RH and LRH non--linear operators, more precisely, those ones contributing to the LH operators after removing the right handed resonances from the physical spectrum and for the hierarchical case $\fL\ll\fR$. From the Tables~\ref{S-T-bounds}-~\ref{hVV-bounds} and the bound in~\eqref{cLR-bound}, we have constrained the following RH and LRH operators
\be
\{\cP_{T,R},\,\cP_{W,R},\,\cP_{1,R},\,\cP_{12,R}\},\qquad\qquad
\{\cP_{C,LR},\,\cP_{T,LR},\,\cP_{W,LR},\,\cP_{\text{3(2)}},\,\cP_{\text{12(1)}},\,\cP_{\text{13(2)}},\,\cP_{\text{17(2)}}\}\,,
\ee

\nt so all in all are the same operators as in~\eqref{Right-operators}-\eqref{Left-Right-operators}, but $\cP_{C,R}$. The latter turns out to be tightly constrained (up to $M^2_W/M^2_{W'}$-corrections), together with $\cP_{C,L}$, by their corresponding contributions to the charged and neutral resonances masses $M_{W^\prime}$ and $M_{Z^\prime}$ respectively in~\eqref{Charged-masses-expanded}-\eqref{Neutral-masses-expanded} as
\be
c_{C,L}\,\,\in\,\,[-4.6,4.6]\times 10^{-5},\,\qquad\qquad
c_{C,R}\,\,\in\,\,[-3.2,3.2]\times 10^{-4}\,.
\label{cL-cR-bounds}
\ee

\boldmath
\subsubsection{Anomalous quartic couplings}
\label{sec:numquartic}
\unboldmath

\nt Among the operators contributing to the quartic gauge couplings in~\eqref{Quartic-gauge-interactions} shown in Table~\eqref{Quartic-gauge-couplings}, some of them were previously bounded either from EWPD constrains, TGC bounds, or limits on $hVV$ couplings. The remaining five operators giving rise to the purely QGC vertices $\{\cP_{6,L},\,\cP_{11,L},\,\cP_{23,L},\,\cP_{24,L},\,\cP_{26,L}\}$ (through their redefined coefficients), are indirectly constraining from their one--loop contribution to the EWPD derived in Ref.~\cite{Brunstein:1996fz}, where it was shown that the five operators correct  $\alpha\Delta T$ while render $\alpha\Delta S=\alpha\Delta U=0$.  In Table~\ref{QGC-bounds} are reported the indirect bounds from Ref.~\cite{Brivio:2013pma} in terms of the implied redefined coefficients, that were determined via the oblique parameters in~(\ref{eq:STUexp}).

\begin{table}[htpb!]
\begin{center}
\begin{tabular}{c||c}
\hline\hline
\\[-4mm]
\bf Coefficients & \bf Ranges (90\% CL)
\\[0.5mm]
\hline\hline
\\[-4mm]
$c_{6,L}$  & $[-0.23, 0.26]$
\\[1mm]
$c_{11,L}$  & $[-0.094, 0.10]$
\\[1mm]
$c_{23,L}$  & $[-0.092, 0.10]$
\\[1mm]
$c_{24,L}$  & $[-0.012, 0.013]$
\\[1mm]
$c_{26,L}$  & $[-0.0061, 0.0068]$
\\[1mm]
\hline\hline
\end{tabular}
\caption{\sf 90\% CL bounds on the anomalous QGC from their 1-loop contribution to the EWPD~\cite{Brivio:2013pma}, assuming only one non-zero operator at a time and for a cutoff $\Lambda_s=2$ TeV.}
\label{QGC-bounds}
\end{center}
\end{table}

\nt On the other hand, anomalous QGC are directly testable at the LHC via three vector bosons production ($VVV$) or in vector boson fusion (VBF) production of two gauge bosons~\cite{Eboli:2000ad}. The best limits come out from the $VVV$--processes at lower center--of--mass energies, whilst the VBF channel dominates for the 14 TeV run~\cite{Eboli:2000ad,Eboli:2003nq,Eboli:2006wa,Belyaev:1998ih,
Degrande:2013yda}. For a 14 TeV at the LHC, the coefficients $\tilde{c}_{6,L}$ and $\tilde{c}_{11,L}$ can be constrained by their combined impact on the VBF channels $p p \to jj W^+ W^-$ and $p p \to jj (W^+ W^+ + W^-W^-)$, where $j$ stands for a tagging jet and the final state $W$'s decay into electron or muon plus neutrino. In~\cite{Eboli:2006wa} were reported the 99\% CL bounds on these coefficients as $ c_{6,L}\,\,\in\,\,[-0.012,\,0.01]$ and  $c_{11,L}\,\,\in\,\,[-0.0077,\,0.014]$, for an integrated luminosity of 100 fb$^{-1}$. These limits are not significantly improved after including the $pp \to jjZZ$ channel~\cite{Belyaev:1998ih}. 

It is worthwhile to comment that for the hypothetical scenario of no LH operators, but only RH and the mixing LRH ones, there will be still contributions to the aforementioned couplings. This is due to the fact that low energy LH operators are effectively turned on once the RH gauge fields are integrated out, feature that is directly reflected in Table~\ref{Redefined-coefficients}.

Notice that no extra operators contribute in the hierarchical case to the coefficients in Table~\ref{QGC-bounds}. Additional terms contribute to them  for a non-small parameter $\epsilon$ and a mixing coefficient $c_{C,LR}\sim \cO(1)$ though. NP effects not far above from the EW scale would point towards a non-small ratio $\epsilon\equiv \fL/\fR$, feature that seems to be favoured by the diboson excess observed at LHC. Such excesses and the impact they entail in our scenario will be analysed in the next.

\section{Comments on the diboson excess at LHC}
\label{Diboson-excess}

\nt Tantalizing deviations from the SM predictions have been recently reported by the ATLAS and CMS Collaborations around invariant mass  of 1.8--2 TeV. They can be interpreted via a $W^\prime$ contribution, and are summarized as:
\begin{itemize}

\item[\bf a)] $3.4\sigma$  local ($2.5\sigma$ global) excess in the ATLAS search~\cite{ATLAS1} (CMS reports a slight excess at the same mass~\cite{CMS-VV}) for a heavy resonance $W^\prime$ decaying as $W^\prime \to WZ \to JJ$, where $J$ stands for two colinear jets from a $W$ or $Z$--boosted decay;

\item[\bf b)] $2.8\sigma$ excess in the CMS search~\cite{CMS}  for a heavy RH boson $W^\prime$ decaying  into an electron and RH neutrino $N$, as $W^\prime \to N\,e \to eejj$; 

\item[\bf c)] a $2.2\sigma$ excess in the CMS search~\cite{CMS1} for $W^\prime\to W h$, with a highly boosted SM Higgs boson $h$ decaying as $h \to b\bar{b}$ and $W\to \ell \nu$ (with $\ell=e,\mu$);

\item[\bf d)] a $2.1\sigma$ excess in the CMS dijet search~\cite{CMS2}.

\end{itemize}
\nt Many scenarios have been proposed in order to account for such excesses. Among them, the left--right EW symmetric model, based on the gauge group $\cG=SU(2)_L\times SU(2)_R\times U(1)_{B-L}$~\cite{LRSM1,LRSM2}, seems to address properly the observed excesses in all the mentioned decay channels. Indeed, the $WZ$ excess (item \textbf{a}) and $W h$ excess (item \textbf{c}) can be tackled~\cite{Dobrescu:2015qna, Gao:2015irw, Brehmer:2015cia} via $W^\prime\to WZ,~W h$, as the implied couplings arise naturally in these models (see~\cite{other} for some alternative explanations of the diboson excess). The $eejj$ excess (item \textbf{b}) can be understood~\cite{Dobrescu:2015qna, Deppisch:2014qpa, Fowlie:2014iua, Gluza:2015goa} through the process $pp\to W^\prime\to N\,e\to eejj$~\cite{KS}, and for a charged gauge boson mass $M_{W^\prime}\sim 2$ TeV, with $g_R<g_L$ at the TeV-scale~\cite{Dobrescu:2015qna}. Finally, the dijet excess (item \textbf{d}) may simply be yielded by $W^\prime\to jj$. 

\nt It is straightforward to translate a $W^\prime$--boson mass nearby 1.8--2 TeV, into our parameter space. Employing again the $W^\prime$ mass formula in~\eqref{Charged-masses-expanded}, we obtain 
\be
\fR\,\approx\,\frac{3.6\text{--}4\,\text{TeV}}{\gR}\,.
\ee

\nt For a coupling $\gR$ in the range $\gR \approx 0.45-0.6$~\cite{Dobrescu:2015qna} we find the $SU(2)_R$ breaking scale $\fR \approx$ 6--8 TeV, entailing thus a ratio $\epsilon\equiv \fL/\fR \approx 0.03-0.04$ for $\fL$ around the EW regime. The small range for $c_{C,LR}$ in~\eqref{cLR-bound} leads then to a negligible parameter $\bar{\epsilon}\sim 10^{-4}$, suppressing as a consequence all the linear and higher $\bar{\epsilon}$--effects induced by the RH and LRH operators onto the LH ones, according with the redefined coefficients in Table~\ref{Redefined-coefficients}. Such effects could be enhanced either via larger strength contributions from the custodial mixing LRH operator $\cP_{C,LR}(h)$, or via NP effects from the right handed gauge sector around the EW scale $\fL$. The former scenario goes against the range in~\eqref{cLR-bound}, spoiling thus the EW value in~\eqref{Redefined-fL}. The second possibility points towards $\epsilon\sim 1$, then $\bar{\epsilon}\sim 0.02$ for a strength contribution of $\cP_{C,LR}(h)$ around its maximal bound in~\eqref{cLR-bound}, and triggering therefore linear $\bar{\epsilon}$--effects. In particular, all the LH operators sector in~\eqref{GT} and the LH basis in~\eqref{CP-even-basis}, will be sensitive to $\bar{\epsilon}$--contributions from the LRH operators, whereas the operators $\cP_{1,L}$ and $\cP_{4,L}$ will have, in addition, extra $\bar{\epsilon}$--terms from some RH operators. This is schematically shown in Table~\ref{Sensitive-operators-comparison}, where the set of operators sensitive to the mentioned effects are collected for both of the $\bar{\epsilon}$--cases.
\begin{table}[htpb!]
\centering
\hspace*{-0.65cm}
\renewcommand{\arraystretch}{1.5}
\begin{tabular}{|c@{\hspace*{0.6mm}}|c|c|} 
\hline 
\multicolumn{2}{|c|}{$\bf \bar{\epsilon}\sim 10^{-4}$} & $\bf \bar{\epsilon}\sim 10^{-2}$\\
\hline\hline
\multicolumn{2}{|c|}{\bf{$\{\cP_B,\,\cP_{C,L},\,\cP_{T,L},\,\cP_{1,L},\,\cP_{2,L},\,\cP_{4,L}\}$}} & \bf{$\{\cP_B,\,\cP_{C,L},\,\cP_{T,L},\,\cP_{W,L},\,\cP_{i,L}\}\quad i=1,...,26$}\\
\hline\hline
\hspace*{-2mm}$\bf R$    & $\{\cP_{C,R},\,\cP_{T,R},\,\cP_{W,R},\,\cP_{1,R},\,\cP_{12,R}\}$ &  $\{\cP_{C,R},\,\cP_{T,R},\,\cP_{W,R},\,\cP_{1,R},\,\cP_{4,R},\,\cP_{12,R},\,\cP_{17,R}\}$ 
\\[2mm] \hline 
\hspace*{-2mm}$\bf LR$    & 
$\begin{array}{l}
\cP_{C,LR},\,\cP_{T,LR},\,\cP_{W,LR},\,\cP_{\text{3(2)}},\,
\\
\phantom{\,\,\,}\cP_{\text{12(1)}},\,\cP_{\text{13(2)}},\,\cP_{\text{17(2)}}
\end{array}$
&
$\begin{array}{l}
\cP_{C,LR},\,\cP_{T,LR},\,\cP_{W,LR},\,\cP_{\text{2(1)}},\,\cP_{\text{3(2-4)}},\,\cP_{\text{5(1-2)}},\,\cP_{\text{6(3)}},\,
\\
\cP_{\text{7-9(1)}},\,\cP_{\text{10(1-2)}},\,\cP_{\text{11(3)}},\,\cP_{\text{12(1)}},\,\cP_{\text{13(2-4)}},\,\cP_{\text{14(1,3,5)}},\,
\cP_{\text{15(1)}},\,
\\
\cP_{\text{16(1,4,6)}},\,\cP_{\text{17(1-2)}},\,\cP_{\text{18(1,3,6)}},\,\cP_{\text{19(1-2)}},\,\cP_{\text{20-22(1)}},\,\cP_{\text{23(3,6)}},\,
\\
\cP_{\text{24(3,6)}},\,\cP_{\text{25(1)}},\,\cP_{\text{26(3)}}
\end{array}$ 
\\[2mm]
\hline 
\end{tabular}
\caption{\sf LH operators (2nd row) sensitive to the RH and LRH operators (3rd and 4th rows) after removing the RH gauge field, and for both of the hierarchical case $\bar{\epsilon}\sim 10^{-4}$ and $\bar{\epsilon}\sim 10^{-2}$ (1st and 2nd big columns). The operators collected here follows the redefined coefficients in Table~\ref{Redefined-coefficients}. A TeV--charged resonances accounting for the observed diboson excess points towards $\bar{\epsilon}\sim 10^{-4}$, and then 5 RH operators and 7 LRH ones contribute to 6 LH operators  (1st big column \& 2nd row). 
NP effects from the right handed gauge sector around the EW scale $\fL$ favours $\epsilon\sim 1$, then $\bar{\epsilon}\sim 0.02$, and consequently 7 RH operators and 43 LRH ones contribute now to 30 LH operators (2nd big column \& 2nd row).} 
\label{Sensitive-operators-comparison}
\end{table}

\nt As it was mentioned before for the bounds on the $hVV$ couplings, the operators set $\{\cP_{7,L},\,\cP_{9,L},\,\cP_{10,L}\}$ was not included since their physical impact on the observables considered is negligible, while the set $\{\cP_{12,L},\,\cP_{17,L},\,\cP_{19,L},\,\cP_{25,L}\}$ entails no relevant contribution for the non-linear realization of the dynamics~\cite{Brivio:2013pma}. In addition, the operators set $\{\cP_{9,L},\,\cP_{10,L},\,\cP_{15,L},\,\cP_{16,L},\,\cP_{19-21,L}\}$ becomes redundant for the massless fermion case via EOM~\cite{Brivio:2013pma}, while the set $\{\cP_{8,L},\,\cP_{18,L},\,\cP_{20-22,L}\}$ does not contribute directly to any of the couplings listed previously. Consequently, when $\bar{\epsilon}$ is not negligible, the additional terms contributing to each one of these operators (LRH according to Table~\ref{Redefined-coefficients}) can also be disregarded. From all these considerations it is concluded that the set of 20 LRH operators
\be
\{\cP_{\text{7-9(1)}},\,\cP_{\text{10(1-2)}},\,\cP_{\text{13(3)}},\,\cP_{\text{15(1)}},\,\cP_{\text{16(1,4,6)}},\,\cP_{\text{17(1)}},\,\cP_{\text{18(1,3,6)}},\,\cP_{\text{19(1-2)}},\,\cP_{\text{20-22(1)}},\,\cP_{\text{25(1)}}\}
\label{LRH-neglected}
\ee

\nt can be disregarded\footnote{Nonetheless the contributions from the $\cP_{\text{12(1)}}$, $\cP_{\text{13(2,4)}}$ and $\cP_{\text{17(2)}}$ enter also through $\cP_{1,L}$, $\cP_{2,L}$ and $\cP_{4,L}$ respectively, being constrained then by the corresponding bounds on the latter LH operators.} for a non-small parameter $\bar{\epsilon}$. By implementing the currents ranges that were obtained from EWPD bounds, TGC limits, $hVV$-couplings constrains and QGC bounds in Tables~\ref{S-T-bounds}-\ref{QGC-bounds} respectively, it is possible to limit as well the additional emerging operators for a non-neglible $\bar{\epsilon}$ via the redefined coefficients in Table~\ref{Redefined-coefficients}. This is shown in Appendix~\ref{App:LH-RH-LRH-Ranges}  Table~\ref{All-RH-LRH-coefficients}, where all the corresponding coefficients ranges for both of the RH or LRH operators are collected and compared with respect to those previously reported for the hierarchical case $\bar{\epsilon}\sim 10^{-4}$. It is possible to conclude from Table~\ref{Sensitive-operators-comparison} and the set in~\eqref{LRH-neglected}, that 7 RH operators and 23 LRH ones can be bounded through the obtained allowed ranges for the case of $\bar{\epsilon}\sim 10^{-2}$. 

The hypothetical scenario of larger strength contributions from the operator $\cP_{C,LR}(h)$ ($\bar{\epsilon} \sim 1$), will drive all the RH and LRH operators in~\eqref{CP-even-basis} and \eqref{CP-even-left-right-basis-I}-\eqref{CP-even-left-right-basis-III} respectively, to contribute onto all the LH operators sector in~\eqref{GT} and the LH basis in~\eqref{CP-even-basis}. The following set of 12 RH \footnote{$\cP_{12,R}$, $\cP_{13,R}$ and $\cP_{17,R}$ enter through $\cP_{1,L}$, $\cP_{2,L}$ and $\cP_{4,L}$ respectively, being then constrained by the corresponding bounds on the latter LH operators.} and 27 LRH operators
\be
\{\cP_{7-10,R},\cP_{15-16,R},\,\cP_{18-22,R},\,\cP_{25,R}\}
\label{RH-neglected-larger}
\ee
\vspace*{-4mm}
\be
\{\cP_{\text{7-9(1)}},\,\cP_{\text{10(1-2)}},\,\cP_{\text{13(1,3)}},\,\cP_{\text{15(1)}},\,\cP_{\text{16(1-6)}},\,\cP_{\text{17(1)}},\,\cP_{\text{18(1-6)}},\,\cP_{\text{19(1-2)}},\,\cP_{\text{20-22(1)}},\,\cP_{\text{25(1)}}\}
\label{RH-LRH-neglected-larger}
\ee

\nt can be disregarded and being possible to constrain therefore 17 RH and 48 LRH operators. The corresponding coefficients ranges for the remaining operators are also  gathered in Appendix~\ref{App:LH-RH-LRH-Ranges}. Some remarks are in order:
\begin{itemize}

\item The ranges for $c_{T,R}$ and $c_{T,LR}$ are unmodified as they are insensitive to additional $\bar{\epsilon}$-corrections (see Table~\ref{Redefined-coefficients}).

\item The corresponding percent and 10\% level ranges for $\{c_{1,R},\,c_{12,R},\,c_{W,LR}\}$ and $\{c_{3(2)},\,c_{13(2)}\}$ respectively, become slightly modified for $\bar{\epsilon}\sim 10^{-2}$ with respect to the hierarchical case, reaching correspondingly a precision of the thousandth and percent level for the hypothetical case $\bar{\epsilon}\sim 1$.

\item The ranges of order $\cO(1-10)$ for $\{c_{4,R},\,c_{17,R}\}$ and $\{c_{5(1-2)},\,c_{6(3)},\,c_{11(3)},\,c_{23(3,6)}\}$ when $\bar{\epsilon}\sim 10^{-2}$, become smaller and around the 10\% level for $\bar{\epsilon}\sim 1$. Similarly, the ranges for $\{c_{2(1)},\,c_{3(3-4)}\}$ turn out to be more precise and constrained at the percent level. 

\item The ranges for $c_{12(1)}$ are all at the thousandth level, slightly modified for $\bar{\epsilon}\sim 10^{-2}$ with respect to the hierarchical case, and becoming half of the range for $\bar{\epsilon}\sim 1$ compared with the one for $\bar{\epsilon}\sim 10^{-4}$.

\end{itemize}

\nt All these remarks lead us therefore to establish the number of the most relevant set of non--linear LH, RH, and LRH operators for the three different $\bar{\epsilon}$--scenarios. Quantitatively one has
\begin{table}[htpb!]
\begin{center}
\begin{tabular}{c|c@{\hspace*{10mm}}c@{\hspace*{10mm}}c@{\hspace*{10mm}}}
& $\bf\bar{\epsilon}\sim 10^{-4}$ & $\bf\bar{\epsilon}\sim 10^{-2}$ & $\bf\bar{\epsilon}\sim 1$
\\[1mm]
\hline
\\[-2mm]
\bf LH  & 16 (13)    &  16 (13)    &  16 (13)
\\
\\
\bf RH  & 5 (4)      &  7 (6)      &  17 (12)
\\
\\
\bf LRH & 7 (5)      &  23 (21)     &  48 (41)
\\
\\[-3mm]
\hline
\\[-2mm]
\bf Total & 28 (22)      &  46 (40)     &  81 (66)
\\
\end{tabular}
\end{center}
\end{table}

\nt In parentheses it has been pointed out the number of remaining operators after neglecting those ones with a corresponding coefficient bounded at the thousandth level. All the remaining operators can be easily identified by keeping track of their corresponding coefficients in Appendix~\ref{App:LH-RH-LRH-Ranges} Table~\ref{All-RH-LRH-coefficients}.


\section{Conclusions}
\label{Conclusions}

\nt In the hypothetical situation of discovering non-zero NP effects at the LHC and future colliders, an effective Lagrangian description would be necessary in order to parametrize all the physical signals detectable at low energies. In this work such NP scenario is pictured by the existence of spin--1 resonances sourced by the extension of the SM local gauge symmetry $\cG_{SM}=SU(2)_L\otimes U(1)_Y$ up to the larger local group $\cG=SU(2)_L\otimes SU(2)_R\otimes U(1)_{B-L}$, here described via a non--linear EW dynamical Higgs scenario, and up to the $p^4$-order in the Lagrangian expansion.

The left-right framework provided here can be considered as a generic UV completion of the low energy non--linear treatment of Refs.~\cite{Appelquist:1980vg,Longhitano:1980iz,Longhitano:1980tm,
Feruglio:1992wf,Appelquist:1993ka}  and  Refs.~\cite{Alonso:2012px,Brivio:2013pma}. Its physical impact has been analysed by integrating out the right handed gauge sector from the physical spectrum, leading the RH and the mixing LRH operators to collapse directly onto the LH sector, and inducing therefore corrections in all the effective pure gauge and gauge-Higgs couplings. These corrections are entirely parametrized in Table~\ref{Redefined-coefficients} via the weighting powers of $\bar{\epsilon}\equiv \epsilon\,c_{C,LR}$, with the scale ratio $\epsilon\equiv \fL/\fR$ and the coefficient $c_{C,LR}$ encoding the strength of the contribution from the mixing operator $\cP_{C,LR}(h)$. This feature leads to modify, consequently, the EWPD parameters (Eqs.~\eqref{S-parameter-hierarchical}-\eqref{T-parameter-hierarchical}), the TGC (Table~\ref{TGV-couplings-BSM}), $hVV$--couplings (Table~\ref{Cubic-gauge-h-couplings}) and the anomalous QGC as well (Table~\ref{Quartic-gauge-couplings}). Corresponding allowed ranges for the involved coefficients have also been reported through Tables~\ref{S-T-bounds}-\ref{QGC-bounds} respectively. In the hypothetical scenario of no LH operators, but only RH and the mixing LRH operators, there will be still contributions to the aforementioned couplings. This is due to the fact that low energy LH operators are effectively turned on once the RH gauge field sector is integrated out, feature that is directly reflected in Table~\ref{Redefined-coefficients}.

The recently observed diboson excess at the ATLAS and CMS Collaborations around the invariant mass of 1.8--2 TeV entails a scale $\fR \sim$ 6--8 TeV, leading to a negligible parameter $\bar{\epsilon}\sim 10^{-4}$ and suppressing therefore all the linear and higher $\bar{\epsilon}$--effects induced by the RH and LRH operators. These effects could be enhanced either via larger strength contributions from the custodial mixing operator $\cP_{C,LR}(h)$, or via NP effects from the right handed gauge sector around the EW scale $\fL$ together with a strength contribution of $\cP_{C,LR}(h)$ around its maximal bound in~\eqref{cLR-bound}. The former scenario spoils the EW value in~\eqref{Redefined-fL}, whereas the latter one points towards $\bar{\epsilon}\sim 10^{-2}$. The set of relevant non-linear LH, RH, and LRH effective operators have been completely identified for the latter regimes and by disregarding: i) irrelevant LH operators with negligible physical impact on the observables considered for the $hVV$--bounds, ii) irrelevant operators for the non-linear realization of the dynamics and redundant for the massless fermion case; iii) operators without any direct contribution to the pure gauge  and gauge--Higgs couplings listed above.

The hypothetical case of larger strength contributions from $\cP_{C,LR}(h)$ would point towards $\bar{\epsilon}\sim 1$, enhancing therefore additional contributions from the RH and LRH operators, and required thus an effective basis of 81 operators in total\,\,=\,\,16\,LH\,\,+\,\,17\,RH\,\,+\,\,48\,LRH (or 66 neglecting irrelevant ones\,\,=\,\,13\,LH\,\,+\,\,12\,RH\,\,+\,\,41\,LRH). The small range for $c_{C,LR}$ leads to $\bar{\epsilon}\sim 10^{-2}$, requiring thus a smaller number of effective operators of 46 operators\,\,=\,\,16\,\,+\,\,7\,\,+\,\,23\,(or 40 without irrelevant ones\,\,=\,\,13\,\,+\,\,6\,\,+\,\,21). The diboson excess around the invariant mass 1.8-2 TeV entails a suppression of $\bar{\epsilon}\sim 10^{-4}$, and therefore the low energy effects will be encoded via a much smaller effective basis with 28 operators in total\,\,=\,\,16\,\,+\,\,5\,\,+\,\,7\,(or 22 relevant\,\,=\,\,13\,\,+\,\,4\,\,+\,\,5). The set of remaining operators are identified by their corresponding coefficients in Table~\ref{All-RH-LRH-coefficients}.  A more detailed interpretation of the diboson excess it is also possible via the left--right non-linear Higgs approach studied here and it can be found in~\cite{Yepes-III}.

\section*{Acknowledgements}

\nt The authors of this work acknowledge valuable and enlightening comments from J.~Gonzalez-Fraile. J.~Y. also acknowledges KITPC financial support during the completion of this work.

 %
%
 \newpage
  
\appendix
\small
%
\section{Operators list}
\label{App:Operators-list}

\boldmath
\subsection{$\cG$--extension of $\LL_0\,+\,\LL_{0,R}$: $\Delta\LL_{\text{CP}}$}
\label{App:Operators-list-L-R}
\unboldmath

\nt The complete linearly independent set of 26 CP--conserving pure gauge and gauge--$h$ non--linear $\cG$--invariant operators up to the $p^4$-order in the effective Lagrangian expansion, and encoded by $\cP_{i,\,L}(h)$ (first term in the second line of $\Delta \LL_{\text{CP},\,L}$, Eq.~\eqref{DeltaL-CP-even-L}) have completely been listed in  Refs.~\cite{Alonso:2012px,Brivio:2013pma}. The symmetric counterpart of $\cP_{i,\,L}(h)$, i.e. $\cP_{i,\,R}(h)$ (second term in $\Delta \LL_{\text{CP},\,R}$ of Eq.~\eqref{DeltaL-CP-even-R}), are 52 non--linear operators in total, among them, 38 (19 $\cP_{i,\,L}(h)$ + 19 $\cP_{i,\,R}(h)$) had already been listed in  Refs.~\cite{Zhang:2007xy,Wang:2008nk}. The whole tower of operators making up the basis $\{\cP_{i,\,L}(h),\,\cP_{i,\,R}(h)\}$ is given by:

\beq
\hspace{-0.8cm}
\begin{aligned}
&\cP_{1,\,\chi} = \gchi\,g' \,B_{\mu\nu}\,\Tr\Big(\TLchi\,\WWchiu\Big)\,\cF_{1,\,\chi}\,,
&&\cP_{14,\,\chi}=\gchi\,\epsilon_{\mu\nu\rho\sigma}\,\Tr\Big(\TLchi\,\VLchimuu\Big)\,\Tr\Big(\VLchinuu\,\WWWchiu\Big)\,\cF_{14,\,\chi}\,, \\[1.4mm] 
&\cP_{2,\,\chi} = i\,g' \,B_{\mu\nu}\,\Tr\Big(\TLchi\,\Big[\VLchimuu,\VLchinuu\Big]\Big)\,\cF_{2,\,\chi}\,,
&&\brown{\cP_{15,\,\chi} = \Big(\Tr\Big(\TLchi\,\cD_\mu\VLchimuu\Big)\Big)^2\,\cF_{15,\,\chi}}\,, \\[1.4mm] 
&\cP_{3,\,\chi}  = i\,\gchi\,\Tr\Big(\WWchiu\,\Big[\VLchimud,\VLchinud\Big]\Big)\,\cF_{3,\,\chi}\,,
&&\brown{\cP_{16,\,\chi} = \Tr\Big(\Big[\TLchi,\VLchinud\Big]\,\cD_\mu \VLchimuu\Big) \, \Tr\Big(\TLchi\,\VLchinuu\Big)\,\cF_{16,\,\chi}}\,,\\[1.4mm]
&\cP_{4}= i\,g'\,B_{\mu\nu}\,\Tr\Big(\TLchi\,\VLchimuu\Big)\,\derp^\nu \cF_{4}\,,
&&\cP_{17,\,\chi} = i\,\gchi \,\Tr\Big(\TLchi\,\WWchiu\Big)\,\Tr\Big(\TLchi\,\VLchimud\Big)\,\derp_\nu \cF_{17,\,\chi}\,, \\[1.4mm]
&\cP_{5,\,\chi} = i\,\gchi \,\Tr\Big(\WWchiu\,\VLchimud\Big)\,\derp_\nu \cF_{5,\,\chi}\,,
&&\cP_{18,\,\chi} = \tr\Big(\TLchi\Big[\VLchimuu,\VLchinuu\Big]\Big)\,\tr\Big(\TLchi\,\VLchimud\Big)\, \derp_\nu\cF_{18,\,\chi}\,,  \\[1.4mm] 
&\cP_{6,\,\chi} =\Big(\Tr\Big(\VLchimud\,\VLchimuu\Big)\Big)^2\,\cF_{6,\,\chi}\,,
&&\brown{\cP_{19,\,\chi} = \Tr\Big(\TLchi\,\cD_\mu\VLchimuu\Big)\,\Tr\Big(\TLchi\,\VLchinuu\Big)\,\derp_\nu \cF_{19,\,\chi}}\,, \\[1.4mm] 
&\brown{\cP_{7,\,\chi} =\Tr\Big(\VLchimud\,\VLchimuu\Big)\,\derp_\nu\derp^\nu\cF_{7,\,\chi}}\,,
&&\cP_{20,\,\chi} = \tr\Big(\VLchimud\,\VLchimuu\Big)\,\derp_\nu\cF_{20,\,\chi}\derp^\nu\cF_{20,\,\chi}'\,, \\[1.4mm]
&\cP_{8,\,\chi} =\Tr\Big(\VLchimuu\,\VLchinuu\Big)\,\derp_\mu\cF_{8,\,\chi}\,\derp_\nu\cF'_{8,\,\chi}\,,
&&\cP_{21,\,\chi} = \Big(\tr\Big(\TLchi\,\VLchimuu\Big)\Big)^2\,\derp_\nu\cF_{21,\,\chi}\,\derp^\nu\cF_{21}'\,, \\[1.4mm] 
&\brown{\cP_{9,\,\chi} = \Tr\Big(\Big(\cD_\mu\VLchimuu\Big)^2 \Big)\,\cF_{9,\,\chi}} \,,
&&\cP_{22,\,\chi} = \Big(\tr\Big(\TLchi\,\VLchimuu\Big)\,\derp_\mu\cF_{22,\,\chi}\Big)^2\,,\\[1.4mm] 
&\brown{\cP_{10,\,\chi} =\Tr\Big(\VLchinuu\,\cD_\mu\VLchimuu\Big)\,\derp_\nu \cF_{10,\,\chi}}\,,
&&\cP_{23,\,\chi} = \tr\Big(\VLchimud\,\VLchimuu\Big)\,\Big(\tr\Big(\TLchi\VLchinuu\Big)\Big)^2\, \cF_{23,\,\chi}\,,\\[1.4mm] 
&\cP_{11,\,\chi} = \Big(\tr\Big(\VLchimuu\,\VLchinuu\Big)\Big)^2\,\cF_{11,\,\chi}\,,
&&\cP_{24,\,\chi} = \tr\Big(\VLchimuu\,\VLchinuu\Big)\,\tr\Big(\TLchi\,\VLchimud\Big)\,\tr\Big(\TLchi\,\VLchinud\Big)\,\cF_{24,\,\chi}\,,\\[1.4mm] 
&\cP_{12,\,\chi} = \gchi^2\,\Big(\Tr\Big(\TLchi\,\WWchiu\Big)\Big)^2\,\cF_{12,\,\chi}\,,
&&\brown{\cP_{25,\,\chi}= \Big(\tr\Big(\TLchi\,\VLchimuu\Big)\Big)^2\,\derp_\nu\derp^\nu\cF_{25,\,\chi}}\,,\\[1.4mm] 
&\cP_{13,\,\chi} = i\,\gchi\,\Tr\Big(\TLchi\,\WWchiu\Big)\,\Tr\Big(\TLchi\,\Big[\VLchimud,\VLchinud\Big]\Big)\,\cF_{13,\,\chi}\,,
&&\cP_{26,\,\chi}=\Big(\Tr\Big(\TLchi\,\VLchimuu\Big)\,\Tr\Big(\TLchi\,\VLchinuu\Big)\Big)^2\,\cF_{26,\,\chi}\,, 
\end{aligned}
\label{CP-even-basis}
\eeq

\nt with $\WWchiu\equiv W^{\mu\nu,a}_\chi\tau^a/2$. In red color have been highlighted all those operators already listed in the context of purely EW chiral effective theories coupled to a light Higgs in Refs.~\cite{Alonso:2012px,Brivio:2013pma} (for $\chi=L$) and not provided in the left-right symmetric EW chiral treatment of Refs.~\cite{Zhang:2007xy,Wang:2008nk}. In Eq.~\eqref{CP-even-basis}, $\DLL_\mu$ denotes the covariant derivative on a field transforming 
in the adjoint representation of $SU(2)_L$, and defined as
\be
\DLL^\mu \VLchinuu \equiv \partial^\mu \VLchinuu\,+\,i\,\gchi\left[ W^\mu_\chi, \VLchinuu \right]\,,\quad \chi=L,R\,.
\label{DV-covariant-derivative}
\ee

\boldmath
\subsection{$SU(2)_L-SU(2)_R$ interplay: $\Delta\LL_{\text{CP},LR}$}
\label{App:Operators-list-LR}
\unboldmath

\nt The local rotations induced by the group $\cG$ are
\be
\LT(x)\equiv e^{\frac{i}{2}\tau^a\alpha^a_L(x)},\,\qquad\qquad  
\RT(x)\equiv e^{\frac{i}{2}\tau^a\alpha^a_R(x)},\, \qquad\qquad
\UY(x)\equiv e^{\frac{i}{2}\tau^3\alpha^0(x)}
\label{Local-transformations}
\ee

\nt with $\alpha^a_{L,R}(x)$ and $\alpha^0(x)$ space-time dependent variables parametrizing the local symmetry transformations, and the Goldstone boson matrices $\UH_{L(R)}$ transforming locally as
\be
\UH_L \rightarrow \mathbf{L}\,\UH_L\,\mathbf{U}^\dagger_Y\,, \qquad\qquad
\UH_R \rightarrow \mathbf{R}\,\UH_R\,\mathbf{U}^\dagger_Y
\label{U-transformation-properties}
\ee

\nt The adjoint vectorial and scalar quantities $\VLLRmuu$ and $\TLLR$ behave covariantly under the transformations in~\eqref{Local-transformations} as
\be
\VLLmuu \rightarrow \quad\LT\,\VLLmuu\,\LT^\dagger\,,\qquad 
\VLRmuu \rightarrow \quad\RT\,\VLRmuu\,\RT^\dagger\,,\qquad \text{and}\qquad  
\TLL \rightarrow \quad\LT\,\TLL\,\LT^\dagger\,,\qquad 
\TLR \rightarrow \quad\RT\,\TLR\,\RT^\dagger\,.
\label{V-T-transformation-properties}
\ee
\nt Therefore, the introduced objects in~\eqref{Vtilde-Ttilde-Wtilde}
will correspondingly behave as
\be
\hspace*{-2.5mm}
\VLLmuut \rightarrow \quad \mathbf{U}_Y\,\VLLmuut\,\mathbf{U}^\dagger_Y\,,\quad
\VLRmuut \rightarrow \quad \mathbf{U}_Y\,\VLRmuut\,\mathbf{U}^\dagger_Y\,, \quad \text{and}\qquad
\TLLt \rightarrow \quad \mathbf{U}^\dagger_Y
\,\TLLt\,\mathbf{U}^\dagger_Y
\,,\quad
\TLRt \rightarrow \quad \mathbf{U}_Y
\,\TLRt\,\mathbf{U}^\dagger_Y
\,,
\label{New-V-T-transformation-properties}
\ee
\nt allowing thus to construct out explicit operators mixing the left--right handed covariant structures $\VLLRmuu$ and $\TLLR$. Similar reasoning applies for the strength gauge fields $\WWLRu$~\cite{Yepes:2015zoa}. In here are listed the operators $\cP_{i(j),LR}(h)$ (second term Eq.~\eqref{DeltaL-CP-even-LR}) where the index $j$ spans over all the possible operators that can be built up from each $\cP_{i,\chi}(h)$ in Eq.~\eqref{CP-even-basis}(with associated coefficients $c_{i(j),LR}$). The complete set of operators $\cP_{i(j),LR}(h)$ listed as:

\beq
\hspace{-0.5cm}
\begin{aligned}
&\cP_{2(1)} = i\,g' \,B_{\mu\nu} \Tr\left(\TLLt\left[\VLLmuut,\VLRnuut\right]\right)\,\cF_{2(1)}\,, \qquad\qquad
&&\brown{\cP_{16(4)} = \Tr\left([\TLRt,\VLRnuut]\,\cD_\mu \VLLmuut\right)\,\Tr\left(\TLLt\,\VLLnudt\right)\,\cF_{16(4)}}\,,\\[2.3mm]
&\cP_{3(1)}  = i\,\gL\,\Tr\left(\WWLut\left[\VLRmudt,\VLRnudt\right]\right)\,\cF_{3(1)}\,,\qquad\qquad
&&\brown{\cP_{16(5)} = \Tr\left([\TLRt,\VLRnuut]\,\cD_\mu \VLLmuut\right)\,\Tr\left(\TLRt\,\VLRnudt\right)\,\cF_{16(5)}}\,,\\[2.3mm]
&\cP_{3(2)}  = i\,\gR\,\Tr\left(\WWRut\left[\VLLmudt,\VLLnudt\right]\right)\,\cF_{3(2)}\,,\qquad\qquad
&&\brown{\cP_{16(6)} = \Tr\left([\TLLt,\VLLnuut]\,\cD_\mu \VLRmuut\right)\,\Tr\left(\TLLt\,\VLLnudt\right)\,\cF_{16(6)}}\,,\\[2.3mm]
&\cP_{3(3)}  = i\,\gL\,\Tr\left(\WWLut\left[\VLLmudt,\VLRnudt\right]\right)\,\cF_{3(3)}\,,\qquad\qquad
&&\cP_{17(1)} = i\,\gL\,\Tr\left(\TLLt\,\WWLut\right)\,\Tr\left(\TLRt\,\VLRmudt\right)\,\derp_\nu \cF_{17(1)}\,, \\[2.3mm]
&\cP_{3(4)}  = i\,\gR\,\Tr\left(\WWRut\left[\VLLmudt,\VLRnudt\right]\right)\,\cF_{3(4)}\,,\qquad\qquad
&&\cP_{17(2)} = i\,\gR\,\Tr\left(\TLRt\,\WWRut\right)\,\Tr\left(\TLLt\,\VLLmudt\right)\,\derp_\nu \cF_{17(2)}\,, \\[2.3mm]
&\brown{\cP_{5(1)} = i\,\gL\,\Tr\left(\WWLut\,\VLRmudt\right)\,\derp_\nu \cF_{5(1)}}\,,\qquad\qquad
&&\cP_{18(1)} = \Tr\left(\TLLt\,[\VLLmuut,\VLLnuut]\right)\,\Tr\left(\TLRt\,\VLRmudt\right)\,\derp_\nu\cF_{18(1)}\,, \\[2.3mm]
&\brown{\cP_{5(2)} = i\,\gR\,\Tr\left(\WWRut\,\VLLmudt\right)\,\derp_\nu \cF_{5(2)}}\,,\qquad\qquad
&&\cP_{18(2)} = \Tr\left(\TLRt\,[\VLRmuut,\VLRnuut]\right)\Tr\left(\TLLt\,\VLLmudt\right)\derp_\nu\cF_{18(2)}\,,\\[2.3mm] 
&\cP_{6(1)} =\left(\Tr\left(\VLLmuut\,\VLRmudt\right)\right)^2\cF_{6(1)}\,,
&&\brown{\cP_{18(3)} = \Tr\left(\TLLt\,[\VLLmuut,\VLRnuut]\right)\Tr\left(\TLLt\,\VLLmudt\right)\derp_\nu\cF_{18(3)}}\,,  
\\[2.3mm]  
&\cP_{6(2)} =\Tr\left(\VLLmuut\,\VLLmudt\right)\,\Tr\left(\VLRnuut\,\VLRnudt\right)\cF_{6(2)}\,,
&&\brown{\cP_{18(4)} = \Tr\left(\TLRt\,[\VLLmuut,\VLRnuut]\right)\Tr\left(\TLRt\,\VLRnudt\right)\derp_\mu\cF_{18(4)}}\,,\\[2.3mm] 
&\cP_{6(3)} =\Tr\left(\VLLmuut\,\VLLmudt\right)\,\Tr\left(\VLLnuut\,\VLRnudt\right)\cF_{6(3)}\,,
&&\brown{\cP_{18(5)} = \Tr\left(\TLLt\,[\VLLmuut,\VLRnuut]\right)\Tr\left(\TLRt\,\VLRmudt\right)\derp_\nu\cF_{18(5)}}\,,\\[2.3mm]  
\end{aligned}
\label{CP-even-left-right-basis-I}
\eeq

\beq
\hspace{-1.5cm}
\begin{aligned}   
&\cP_{6(4)} =\Tr\left(\VLRmuut\,\VLRmudt\right)\,\Tr\left(\VLLnuut\,\VLRnudt\right)\cF_{6(4)}\,,
&&\brown{\cP_{18(6)} = \Tr\left(\TLRt\,[\VLLmuut,\VLRnuut]\right)\Tr\left(\TLLt\,\VLLnudt\right)\derp_\mu\cF_{18(6)}}\,,\\[2.3mm]
&\brown{\cP_{7(1)} =\Tr\left(\VLLmuut\,\VLRmudt\right)\,\derp_\nu\derp^\nu\cF_{7(1)}}\,,
&&\brown{\cP_{19(1)} = \Tr\left(\TLLt\,\cD_\mu\VLLmuut\right)\Tr\left(\TLRt\,\VLRnuut\right)\derp_\nu \cF_{19(1)}}\,, 
\\[2.3mm]
&\brown{\cP_{8(1)} =\Tr\left(\VLLmuut\,\VLRnuut\right)\,\derp_\mu\cF_{8(1)}\,\derp_\nu\cF'_{8(1)}}\,,
&&\brown{\cP_{19(2)} = \Tr\left(\TLRt\,\cD_\mu\VLRmuut\right)\Tr\left(\TLLt\,\VLLnuut\right)\derp_\nu \cF_{19(2)}}\,, 
\\[2.3mm]
&\brown{\cP_{9(1)} = \Tr\left(\cD_\mu\VLLmuut\,\cD_\nu\VLRnuut \right)\,\cF_{9(1)}} \,,
&&\brown{\cP_{20(1)} = \Tr\left(\VLLmudt\,\VLRmuut\right)\,\derp_\nu\cF_{20(1)}\,\derp^\nu\cF'_{20(1)}}\,,\\[2.3mm]
&\brown{\cP_{10(1)} =\Tr\left(\VLLnuut \,\cD_\mu\VLRmuut\right)\,\derp_\nu \cF_{10(1)}}\,,
&&\cP_{21(1)} = \Tr\left(\TLLt\,\VLLmuut\right)\Tr\left(\TLRt\,\VLRmudt\right)\left(\derp_\nu\cF_{21(1)}\right)^2\,,\\[2.3mm]
&\brown{\cP_{10(2)} =\Tr\left(\VLRnuut \,\cD_\mu\VLLmuut\right)\,\derp_\nu \cF_{10(2)}}\,,
&&\cP_{22(1)} = \Tr\left(\TLLt\,\VLLmuut\right)\Tr\left(\TLRt\,\VLRnuut\right)\derp_\mu\cF_{22(1)}\derp_\nu\cF'_{22(1)}\,,\\[2.3mm]
&\cP_{11(1)} = \left(\Tr\left(\VLLmuut\,\VLRnuut\right)\right)^2\,\cF_{11(1)}\,,
&&\cP_{23(1)} = \Tr\left(\VLLmuut\,\VLLmudt\right)\,\left(\Tr\left(\TLRt\,\VLRnuut\right)\right)^2\, \cF_{23(1)}\,,\\[2.3mm]
&\cP_{11(2)} = \Tr\left(\VLLmuut\,\VLLnuut\right)\Tr\left(\VLRmudt\,\VLRnudt\right)\,\cF_{11(2)}\,,
&&\cP_{23(2)} = \Tr\left(\VLRmuut\,\VLRmudt\right)\,\left(\Tr\left(\TLLt\,\VLLnuut\right)\right)^2\, \cF_{23(2)}\,,\\[2.3mm]
&\cP_{11(3)} = \Tr\left(\VLLmuut\,\VLLnuut\right)\Tr\left(\VLLmudt\,\VLRnudt\right)\,\cF_{11(3)}\,,
&&\cP_{23(3)} = \Tr\left(\VLLmuut\,\VLLmudt\right)
\Tr\left(\TLLt\,\VLLnuut\right)\Tr\left(\TLRt\,\VLRnudt\right) \cF_{23(3)}\,,
\\[2.3mm]
&\cP_{11(4)} = \Tr\left(\VLRmuut\,\VLRnuut\right)\Tr\left(\VLLmudt\,\VLRnudt\right)\,\cF_{11(4)}\,,
&&\cP_{23(4)} = \Tr\left(\VLRmuut\,\VLRmudt\right)
\Tr\left(\TLLt\,\VLLnuut\right)\Tr\left(\TLRt\,\VLRnudt\right) \cF_{23(4)}\,,\\[2.3mm]
&\cP_{11(5)} = \Tr\left(\VLLmuut\,\VLRnuut\right)\Tr\left(\VLRmudt\,\VLLnudt\right)\,\cF_{11(5)}\,,
&&\cP_{23(5)} = \Tr\left(\VLLmuut\,\VLRmudt\right)\,\left(\Tr\left(\TLRt\,\VLRnuut\right)\right)^2\, \cF_{23(5)}\,,\\[2.3mm]
&\cP_{12(1)} = \gL\,\gR\,\Tr\left(\TLLt\,\WWLut\right)\Tr\left(\TLRt\,\WWRdt\right)\,\cF_{12(1)}\,, 
&&\cP_{23(6)} = \Tr\left(\VLLmuut\,\VLRmudt\right)\left(\Tr\left(\TLLt\,\VLLnuut\right)\right)^2 \cF_{23(6)}\,,\\[2.3mm] 
&\cP_{13(1)} = i\,\gL\,\Tr\left(\TLLt\,\WWLut\right)\,\Tr\left(\TLRt\left[\VLRmudt,\VLRnudt\right]\right)\,\cF_{13(1)}\,,
&&\cP_{23(7)} = \Tr\left(\VLLmuut\,\VLRmudt\right)
\Tr\left(\TLLt\,\VLLnuut\right)\Tr\left(\TLRt\,\VLRnudt\right) \cF_{23(7)}\,,\\[2.3mm] 
&\cP_{13(2)} = i\,\gR\,\Tr\left(\TLRt\,\WWRut\right)\,\Tr\left(\TLLt\left[\VLLmudt,\VLLnudt\right]\right)\,\cF_{13(2)}\,,
&&\cP_{24(1)} = \Tr\left(\VLLmuut\,\VLLnuut\right)\Tr\left(\TLRt\,\VLRmudt\right)\Tr\left(\TLRt\,\VLRnudt\right)\cF_{24(1)}\,,\\[2.3mm] 
&\cP_{13(3)} = i\,\gL\,\Tr\left(\TLLt\,\WWLut\right)\,\Tr\left(\TLLt\left[\VLLmudt,\VLRnudt\right]\right)\,\cF_{13(3)}\,,
&&\cP_{24(2)} = \Tr\left(\VLRmuut\,\VLRnuut\right)\Tr\left(\TLLt\,\VLLmudt\right)\Tr\left(\TLLt\,\VLLnudt\right)\cF_{24(2)}\,,\\[2.3mm] 
&\cP_{13(4)} = i\,\gR\,\Tr\left(\TLRt\,\WWRut\right)\,\Tr\left(\TLRt\left[\VLLmudt,\VLRnudt\right]\right)\,\cF_{13(4)}\,,
&&\cP_{24(3)} = \Tr\left(\VLLmuut\,\VLLnuut\right)\Tr\left(\TLLt\,\VLLmudt\right)\Tr\left(\TLRt\,\VLRnudt\right)\cF_{24(3)}\,,\\[2.3mm] 
&\cP_{14(1)} = \gL\,\epsilon_{\mu\nu\rho\sigma}\,\Tr\left(\TLRt\VLRmuut\right)
\Tr\left(\VLLnuut\,\WWWLut\right)\,\cF_{14(1)}\,, 
&&\cP_{24(4)} = \Tr\left(\VLRmuut\,\VLRnuut\right)\Tr\left(\TLLt\,\VLLmudt\right)\Tr\left(\TLRt\,\VLRnudt\right)\cF_{24(4)}\,,\\[2.3mm] 
&\cP_{14(2)} = \gR\,\epsilon_{\mu\nu\rho\sigma}\,\Tr\left(\TLLt\VLLmuut\right)
\Tr\left(\VLRnuut\,\WWWRut\right)\,\cF_{14(2)}\,, 
&&\cP_{24(5)} = \Tr\left(\VLLmuut\,\VLRnuut\right)\Tr\left(\TLRt\,\VLRmudt\right)\Tr\left(\TLRt\,\VLRnudt\right)\cF_{24(5)}\,,\\[2.3mm] 
&\cP_{14(3)} = \gL\,\epsilon_{\mu\nu\rho\sigma}\,\Tr\left(\TLLt\VLLmuut\right)
\Tr\left(\VLRnuut\,\WWWLut\right)\,\cF_{14(3)}\,, 
&&\cP_{24(6)} = \Tr\left(\VLLmuut\,\VLRnuut\right)\Tr\left(\TLLt\,\VLLmudt\right)\Tr\left(\TLLt\,\VLLnudt\right)\cF_{24(6)}\,,\\[2.3mm]  
&\cP_{14(4)} = \gR\,\epsilon_{\mu\nu\rho\sigma}\,\Tr\left(\TLRt\VLRmuut\right)
\Tr\left(\VLLnuut\,\WWWRut\right)\,\cF_{14(4)}\,,
&&\cP_{24(7)} = \Tr\left(\VLLmuut\,\VLRnuut\right)\Tr\left(\TLLt\,\VLLmudt\right)\Tr\left(\TLRt\,\VLRnudt\right)\cF_{24(7)}\,,\\[2.3mm] 
&\cP_{14(5)} = \gR\,\epsilon_{\mu\nu\rho\sigma}\,\Tr\left(\TLLt\VLLmuut\right)
\Tr\left(\VLLnuut\,\WWWRut\right)\,\cF_{14(5)}\,, 
&&\cP_{24(8)} = \Tr\left(\VLLmuut\,\VLRnuut\right)\Tr\left(\TLRt\,\VLRmudt\right)\Tr\left(\TLLt\,\VLLnudt\right)\cF_{24(8)}\,,\\[2.3mm]  
&\cP_{14(6)} = \gL\,\epsilon_{\mu\nu\rho\sigma}\,\Tr\left(\TLRt\VLRmuut\right)
\Tr\left(\VLRnuut\,\WWWLut\right)\,\cF_{14(6)}\,, 
&&\brown{\cP_{25(1)}= \Tr\left(\TLLt\,\VLLmuut\right)\Tr\left(\TLRt\,\VLRmudt\right)\derp_\nu\derp^\nu\cF_{25(1)}}\,,\\[2.3mm]
\end{aligned}
\label{CP-even-left-right-basis-II}
\eeq

\beq
\hspace{-1.5cm}
\begin{aligned}   
&\brown{\cP_{15(1)} = \Tr\left(\TLLt\,\cD_\mu\VLLmuut\right)\,\,\Tr\left(\TLRt\,\cD_\nu\VLRnuut\right)\,\cF_{15(1)}}\,, 
&&\cP_{26(1)}=\left(\Tr\left(\TLLt\,\VLLmuut\right)\Tr\left(\TLRt\,\VLRmudt\right)\right)^2\cF_{26(1)}\,,\\[2.3mm] 
&\brown{\cP_{16(1)} = \Tr\left([\TLLt,\VLLnuut]\,\cD_\mu \VLLmuut\right)\,\Tr\left(\TLRt\,\VLRnudt\right)\,\cF_{16(1)}}\,,
&&\cP_{26(2)}=\left(\Tr\left(\TLLt\,\VLLmuut\right)\Tr\left(\TLRt\,\VLRnuut\right)\right)^2\cF_{26(2)}\,,\\[2.3mm] 
&\brown{\cP_{16(2)} = \Tr\left([\TLRt,\VLRnuut]\,\cD_\mu \VLRmuut\right)\,\Tr\left(\TLLt\,\VLLnudt\right)\,\cF_{16(2)}}\,,
&&\cP_{26(3)}=\Tr\left(\TLLt\,\VLLmuut\right)\Tr\left(\TLRt\,\VLRmudt\right)\left(\Tr\left(\TLLt\,\VLLnuut\right)\right)^2\cF_{26(3)}\,,\\[2.3mm] 
&\brown{\cP_{16(3)} = \Tr\left([\TLLt,\VLLnuut]\,\cD_\mu \VLRmuut\right)\,\Tr\left(\TLRt\,\VLRnudt\right)\,\cF_{16(3)}}\,,
&&\cP_{26(4)}=\Tr\left(\TLLt\,\VLLmuut\right)\Tr\left(\TLRt\,\VLRmudt\right)\left(\Tr\left(\TLRt\,\VLRnuut\right)\right)^2\cF_{26(4)}\,,\\[2.3mm]
\end{aligned}
\label{CP-even-left-right-basis-III}
\eeq

\nt with $\cD_\mu \VLLRmuut$ following a similar definition as in~\eqref{Vtilde-Ttilde-Wtilde}
\be
\cD_\mu \VLLRmuut\equiv \UHLR^\dagger\,\cD_\mu \VLLRmuu\,\UHLR\,,
\label{DVtilde-covariant-derivative}
\ee

\nt and the explicit light Higgs dependence is implicitly assumed through all $\cF_i$ and $\cP_{i(j),LR}$ . Suffix $LR$ in all $\cP_{i(j),LR}(h)$ and their corresponding $\cF_{i(j),LR}(h)$ has been omitted as well in Eqs.~\eqref{CP-even-left-right-basis-I}-\eqref{CP-even-left-right-basis-II}. Among the total 75 operators $\cP_{i(j)}$ listed in Eqs.~\eqref{WT-LR}-\eqref{CP-even-left-right-basis-II}, 23 operators (highlighted in red color again) are missing in the left-right symmetric EW chiral treatment of Refs.~\cite{Zhang:2007xy,Wang:2008nk}.

%
\section{Equations of motion}
\label{App:EOM}

\nt Some of the CP--conserving bosonic operators provided above can be directly traded by pure bosonic operators  plus fermionic-bosonic ones~\cite{Alonso:2012px,Brivio:2013pma,Yepes:2015zoa}. Such connection is established through the covariant derivative $\cD_\mu\,\VLchimuu$ and the corresponding equation of motion for the light Higgs. When fermion masses are neglected, operators containing $\cD_\mu\,\VLchimuu$ can be written in terms of the other operators in the basis $\Delta\LL_{\text{CP}}$ and $\Delta\LL_{\text{CP},LR}$. Additionally, via light $h$--EOM, operators with two derivative couplings of $\cF(h)$ can be reduced to a combination of bosonic operators plus fermionic-bosonic ones. In general, all those operators must be included to have a complete independent bosonic basis.

Considering the LO Lagrangian $\LL_0\,+\,\LL_{0,R}$ described along Eqs.~\eqref{LLO}-\eqref{LLO-Right}, and accounting for the mixing effects from the LRH operators $\cP_{C,LR}(h)$ and $\cP_{W,LR}(h)$ in~\eqref{WT-LR} (the effect of $\cP_{T,LR}(h)$ can be neglected), the EOM for the strength gauge fields $W^{\mu,a}_{L(R)}$ and $B^\mu$, and for the light Higgs $h$, are correspondingly 
\beq
\begin{aligned}
&\Big(D_\mu\,\WWLRu\Big)^a +c_{W,LR}\,\partial_\mu \Big[\Tr\left(\UHLR^\dagger\,\tau^a\,\UHLR\,\WWRLut\right)\Big]  +\frac{\delta\LL_{f-kinetic}}{\delta W^{a}_{\nu,L(R)}}=\\ \\
&  \frac{i}{4}\,\gLR\,\Big\{\fLR^2\,\Tr\left(\VLLRnuu\,\tau^a\right) +c_{C,\,LR}\,\fL\fR\Tr\left(\UHLR^\dagger\,\tau^a\,\UHLR\,\VLRLnuut\right)\Big\}
\left(1+\frac{h}{\fL}\right)^2\\[-2mm]
\label{W-EOM}
\end{aligned}
\eeq

\beq
\begin{aligned}
&\derp_\mu B^{\mu\nu} + \frac{\delta\LL_{f-kinetic}}{\delta B_{\nu}}=\\ \\
& -\frac{i}{4}\,g'\,\Big\{\Big[\fL^2\,\Tr\Big(\TLL\,\VLLnuu\Big)+\fR^2\,\Tr\Big(\TLR\,\VLRnuu\Big)\Big] 
 +\,\fL\,\fR\,c_{C,\,LR}\,\Big[\Tr\Big(\TLRt\,\VLLnuut\Big)+\Tr\Big(\TLLt\,\VLRnuut\Big)\Big]\Big\}\left(1+\frac{h}{\fL}\right)^2
\label{B-EOM}
\end{aligned}
\eeq

\beq
\begin{aligned}
&\square h + \dfrac{\delta V(h)}{\delta h}+\frac{\delta\LL_{Yukawa}}{\delta h}=\\ \\
&-\frac{1}{2\,\fL}\left[\fL^2\,\Tr\Big(\VLLmuu\,\VLLmud\Big)+\fR^2\,\Tr\Big(\VLRmuu\,\VLRmud\Big)\right]\left(1+\frac{h}{\fL}\right)\,\,+\,\,c_{C,\,LR}\,\frac{\fR}{2}\,\Tr\Big(\VLLmuut\,\VLRmudt\Big)\left(1+\frac{h}{\fL}\right)\,,
\label{h-EOM}
\end{aligned}
\eeq

\nt where the fermion dependent part of the EOMs has been generically encoded in the third, second, and third terms at the first lines of Eqs.~\eqref{W-EOM}-\eqref{B-EOM}-\eqref{h-EOM} respectively, as no explicit kinetic fermion terms nor Yukawa interactions were accounted by $\LL_\text{chiral}$ in~\eqref{Lchiral}. From Eqs.~\eqref{W-EOM}-\eqref{B-EOM} it is derived
\beq
\begin{aligned}
&\left[\fR\left(\fR + c_{C,\,LR}\,\fL\right)\Tr\Big(\TLR\,\cD_\mu\VLRmuu\Big)+\left(R \leftrightarrow L\right)\right]\left(1+\frac{h}{\fL}\right)^2=\\ \\
&-2\,\left[\fL\left(\fL + c_{C,\,LR}\,\fR\right)\Tr\Big(\TLL\,\VLLmuu\Big)+\left(L \leftrightarrow R\right)\right]\,\frac{\partial_\mu h}{\fL}\left(1+\frac{h}{\fL}\right)\,\,+\,\,\frac{4\,i}{g'}\,\partial_\mu\Big(\frac{\delta\LL_{f-kinetic}}{\delta B_{\mu}}\Big)\,,\\ 
\end{aligned}
\label{T-DV-R}
\eeq

\beq
\begin{aligned}
&\Big\{\fLR\,\Tr\Big(\tau^a\,\cD_\mu\VLLRmuu\Big) +c_{C,\,LR}\,\fRL\,\Tr\left(\UHLR^\dagger\,\tau^a\,\UHLR\,\cD_\mu \VLRLmuut\right)\Big\}
\left(1+\frac{h}{\fL}\right)^2=\\ \\
&\Big\{-2\,\fLR\Tr\Big(\tau^a\,\VLLRmuu\Big)\,\frac{\partial_\mu h}{\fL} -c_{C,\,LR}\, \fRL\,\Tr\left(\UHLR^\dagger\,\tau^a\,\UHLR\,\left[\VLLRmuut,\VLRLmudt\right]\right)\Big\}
\left(1+\frac{h}{\fL}\right) +\\ \\
&- 2\,c_{C,\,LR}\,\fRL\,\Tr\left(\UHLR^\dagger\,\tau^a\,\UHLR\,\VLRLmuut\right)\,
\frac{\partial_\mu h}{\fL}\,\left(1+\frac{h}{\fL}\right) -\frac{4\,i}{\gLR\,\fLR}\,\partial_\mu\Big(\frac{\delta\LL_{f-kinetic}}{\delta W^{a}_{\mu,L(R)}}\Big)\,,\\ 
\end{aligned}
\label{tau-DV-R}
\eeq

\nt where the last terms in Eqs.~\eqref{T-DV-R}-\eqref{tau-DV-R} can be translated into Yukawa terms via implementation of the corresponding Dirac equations. As it can be seen from the relations above, operators containing the contraction $\cD_\mu\,\VLRmuu$ can be translated into fermionic-bosonic operators plus pure bosonic ones, with some of them containing the contraction $\cD_\mu\,\VLLmuu$ (from the second terms in the left hand side of Eqs.~\eqref{T-DV-R}-\eqref{tau-DV-R}). Furthermore, by using the light Higgs-EOM in Eq.~\eqref{h-EOM}, those operators with two derivative couplings of $\cF(h)$ can be also rewritten in terms of pure bosonic ones plus fermionic-bosonic ones.


\section{(De)correlation formulae}
\label{App:De-correlation-formulae}

\nt Some of the non--linear operators in~\eqref{DeltaL-CP-even-L}, \eqref{DeltaL-CP-even-R} and~\eqref{DeltaL-CP-even-LR} contribute to more than one of the couplings in Tables~\ref{TGV-couplings-BSM}-\ref{Cubic-gauge-h-couplings}. Therefore, a set of relations relating different couplings are possible. For the TGV sector one has
\begin{align}
\Delta\kappa_Z+\frac{s_W^2}{c_W^2}\Delta \kappa_\gamma
-\Delta g_1^Z&=-\frac{8 e^2}{s_W^2}\left(2 \tilde{c}_{12,L}+\tilde{c}_{13,L}\right)\,,
\label{Dec-1}\\
\Delta g_{6}^\gamma+\dfrac{c_W^2}{s_W^2}\Delta g_{6}^Z&=\frac{e^2}{s_W^4}\,\tilde{c}_{16,L}\,,
\label{Dec-2}
\end{align}

\nt while other examples of relations involving $hVV$--couplings are
\begin{align}
2\,c_W^2\,\,g^{(1)}_{hZZ}+2\,s_W\,c_W\,\,g^{(1)}_{\gamma hZ}+
2\,s_W^2\,g_{\gamma\gamma h}-g^{(1)}_{hWW}&
=\frac{4 e^2}{s_W^2}\, \tilde{a}_{12,L}\,,
\label{Dec-3}\\
g^{(2)}_{hWW}-c_W^2\, g^{(2)}_{hZZ}-s_W\,c_W\, g^{(2)}_{\gamma hZ}
&=\frac{2 e^2}{s_W^2}\, \tilde{a}_{17,L}\,,
\label{Dec-4}\\
\Delta g^{(3)}_{hZZ}-\dfrac{1}{c_W^2}\Delta g^{(3)}_{hWW}&
=-\frac{8 e^2}{s_{2 w}^2}\,\tilde{a}_{19,L}\,,
\label{Dec-5}\\
\Delta g^{(4)}_{hZZ}-\dfrac{1}{2\,c_W^2}\Delta g^{(4)}_{hWW}&
=-\frac{8 e^2}{s_{2 w}^2}\,\tilde{a}_{15,L}\,,
\label{Dec-6}\\
\Delta g^{(6)}_{hZZ}-\dfrac{1}{2\,c_W^2}\Delta g^{(6)}_{hWW}&
=-\frac{8 e^2}{s_{2 w}^2}\, \tilde{a}_{25,L}\,.
\label{Dec-7}
\end{align}

\nt where the variations in the left hand side of~(\ref{Dec-1})-(\ref{Dec-2}) and~(\ref{Dec-5})-(\ref{Dec-7}) are corresponding in this scenario to the sum of the values at the 3rd and 4th columns in Tables~\ref{TGV-couplings-BSM}-\ref{Cubic-gauge-h-couplings}. The induced effects encoded through $\alpha_{WB}$ cancels out in~(\ref{Dec-1}). By linking non--linear approaches to the linear effective scenarios explicitly implementing the SM Higgs doublet~\cite{Alonso:2012px,Brivio:2013pma}, and via powers of the weighting parameter $v^2/\fL^2$, it is realized that the right-hand side of  Eqs.~(\ref{Dec-1})-(\ref{Dec-7}) correspond to $v^4/\fL^4$-weighted terms in  the non-linear Lagrangian. They would vanish if: a) the $d=6$ linear limit\footnote{Eq.~(\ref{Dec-1})
with vanishing right-hand side was already
known to hold in the linear
regime at $d=6$~\cite{Hagiwara:1993ck,Hagiwara:1995vp}.}; ii) in the $v^2/\fL^2-$truncated non-linear
Lagrangian; iii) in the custodial preserving limit. The first two relations with a vanishing right-hand side where already found in Ref.~\cite{Contino:2013kra}. Non-zero deviations from zero in the data combinations indicated by the left-hand side of those equations would point towards either $d=8$ corrections of the linear expansion or a non-linear realisation of the underlying dynamics.


\section{Operator coefficients bounds}
\label{App:LH-RH-LRH-Ranges}

\nt By implementing the currents ranges that were obtained from EWPD bounds, TGC limits, $hVV$-couplings constrains and QGC bounds in Tables~\ref{S-T-bounds}-\ref{QGC-bounds} respectively, it is possible to limit as well the additional emerging operators for a non-neglible $\bar{\epsilon}$ via the redefined coefficients in Table~\ref{Redefined-coefficients}. Table~\ref{All-RH-LRH-coefficients} collects all the corresponding coefficients ranges for both of the RH and LRH operators.

\begin{table}[htpb!]
\vspace*{-1.cm}
\hspace*{-2.1cm}
\renewcommand{\arraystretch}{0.85}
\small{
\begin{tabular}{c||c@{\hspace*{0.7mm}}c@{\hspace*{0.7mm}}c@{\hspace*{0.7mm}}||c||c@{\hspace*{0.7mm}}c@{\hspace*{0.7mm}}c@{\hspace*{0.7mm}}}
\hline\hline
\\[-4mm]
\bf Coeff.  & $\bf\bar{\epsilon}\sim 10^{-4}$ & $\bf\bar{\epsilon}\sim 10^{-2}$ & $\bf\bar{\epsilon}\sim 1$ & \bf Coeff.  & $\bf\bar{\epsilon}\sim 10^{-4}$ & $\bf\bar{\epsilon}\sim 10^{-2}$ & $\bf\bar{\epsilon}\sim 1$
\\[0.5mm]
\hline\hline
\\[-3mm]
$c_{C,R}$  & $[-3.2,\,3.2]\times 10^{-4}$ & $[-3.2,\,3.2]\times 10^{-4}$ & $[-3.2,\,3.2]\times 10^{-4}$ & $c_{11,R}$  & $-$ & $-$ & $[-0.094, 0.1]$ 
\\[2mm]
$c_{T,R}$  & $[-0.002,\,0.0017]$ & $[-0.002,\,0.0017]$ & $[-0.002,\,0.0017]$ & $c_{12,R}$  & $[-0.053, 0.13]$ & $[-0.049, 0.12]$ & $[-0.001, 0.001]$
\\[2mm]
$c_{W,R}$  & $[-0.50, 0.21]$ & $[-0.47, 0.21]$ & $[-0.0047, 0.004]$  & 
$c_{13,R}$  & $-$ & $-$ & $[-0.06, 0.038]$
\\[2mm]
$c_{1,R}$  & $[-0.053, 0.13]$ & $[-0.05, 0.12]$ & $[-0.004, 0.0047]$ & 
$c_{14,R}$  & $-$ & $-$ & $[-0.02, 0.04]$ 
\\[2mm]
$c_{2,R}$  & $-$ & $-$ & $[-0.12, 0.076]$ & $c_{17,R}$  & $-$ & $[-4, 13.4]$ & $[-0.14, 0.47]$
\\[2mm]
$c_{3,R}$  & $-$ & $-$ & $[-0.079, 0.064]$ & $c_{23,R}$  & $-$ & $-$ & $[-0.092, 0.1]$
\\[2mm]
$c_{4,R}$  & $-$ & $[-4, 13.4]$ & $[-0.14, 0.47]$ & $c_{24,R}$ & $-$ & $-$ & $[-0.012, 0.013]$
\\[2mm]
$c_{5,R}$  & $-$ & $-$ & $[-0.33, 0.17]$ & $c_{26,R}$  & $-$ & $-$ & $[-0.006, 0.007]$
\\[2mm]
$c_{6,R}$  & $-$ & $-$ & $[-0.23, 0.26]$ & & & & 
\\[1mm]
\hline\hline
\\[-3mm]
$c_{C,LR}$  & $[-0.02,\,0.02]$ & $[-0.02,\,0.02]$ & $\sim 1$ & $c_{14(3)}$ & $-$ & $[-0.58,\,1.1]$ & $[-0.02,\,0.04]$
\\[2mm]
$c_{T,LR}$  & $[-0.0008,\,0.001]$ & $[-0.0008,\,0.001]$ & $[-0.0008,\,0.001]$ & $c_{14(4)}$ & $-$ & $-$ & $[-0.04,\,0.02]$
\\[2mm]
$c_{W,LR}$  & $[-0.016,\,0.018]$ & $[-0.015,\,0.018]$ & $[-0.008,\,0.009]$ & $c_{14(5)}$  & $-$ & $[-0.58,\,1.1]$ & $[-0.02,\,0.04]$
\\[2mm]
$c_{2(1)}$  & $-$ & $[-2.2, 3.4]$ & $[-0.076,\,0.012]$ & $c_{14(6)}$ & $-$ & $-$ & $[-0.04,\,0.02]$ 
\\[2mm]
$c_{3(1)}$  & $-$ & $-$ & $[-0.064,\,0.079]$ & $c_{17(2)}$  & $[-0.47, 0.14]$ & $[-0.45, 0.13]$ & $[-0.24,\,0.07]$
\\[2mm]
$c_{3(2)}$  & $[-0.24,\,0.152]$ & $[-0.23,\,0.14]$ & $[-0.079,\,0.064]$ & $c_{23(1)}$ & $-$ & $-$ & $[-0.09,\,0.1]$
\\[2mm]
$c_{3(3)}$  & $-$ & $[-2.2,\,1.8]$ & $[-0.079,\,0.064]$ & $c_{23(2)}$ & $-$ & $-$ & $[-0.09,\,0.1]$
\\[2mm]
$c_{3(4)}$  & $-$ & $[-4.3,\,6.8]$ & $[-0.064,\,0.079]$ & $c_{23(3)}$  & $-$ & $[-2.8,\,2.6]$ & $[-0.1,\,0.09]$
\\[2mm]
$c_{5(1)}$  & $-$ & $[-4.8,\,9.4]$ & $[-0.17,\,0.33]$ & 
$c_{23(4)}$ & $-$ & $-$ & $[-0.1,\,0.09]$
\\[2mm]
$c_{5(2)}$  & $-$ & $[-4.8,\,9.4]$ & $[-0.17,\,0.33]$ & $c_{23(5)}$ & $-$ & $-$ & $[-0.1,\,0.09]$
\\[2mm]
$c_{6(1)}$  & $-$ & $-$ & $[-0.23,\,0.26]$ & $c_{23(6)}$  & $-$ & $[-2.8,\,2.6]$ & $[-0.1,\,0.09]$
\\[2mm]
$c_{6(2)}$  & $-$ & $-$ & $[-0.23,\,0.26]$ & $c_{23(7)}$ & $-$ & $-$ & $[-0.09,\,0.1]$
\\[2mm]
$c_{6(3)}$  & $-$ & $[-7.4,\,6.6]$ & $[-0.26,\,0.23]$ & $c_{24(1)}$ & $-$ & $-$ & $[-0.012,\,0.013]$
\\[2mm]
$c_{6(4)}$  & $-$ & $-$ & $[-0.26,\,0.23]$ & $c_{24(2)}$ & $-$ & $-$ & $[-0.012,\,0.013]$
\\[2mm]
$c_{11(1)}$ & $-$ & $-$ & $[-0.094,\,0.1]$ & $c_{24(3)}$ & $-$ & $[-0.37,\,0.34]$ & $[-0.013,\,0.012]$
\\[2mm]
$c_{11(2)}$ & $-$ & $-$ & $[-0.094,\,0.1]$ & $c_{24(4)}$ & $-$ & $-$ & $[-0.013,\,0.012]$
\\[2mm]
$c_{11(3)}$ & $-$ & $[-2.8,\,2.7]$ & $[-0.1,\,0.094]$ & $c_{24(5)}$ & $-$ & $-$ & $[-0.013,\,0.012]$
\\[2mm]
$c_{11(4)}$ & $-$ & $-$ & $[-0.1,\,0.094]$ & $c_{24(6)}$ & $-$ & $[-0.37,\,0.34]$ & $[-0.013,\,0.012]$
\\[2mm]
$c_{11(5)}$ & $-$ & $-$ & $[-0.094,\,0.1]$ & $c_{24(7)}$ & $-$ & $-$ & $[-0.012,\,0.013]$
\\[2mm]
$c_{12(1)}$ & $[-0.0047,\,0.004]$ & $[-0.0045,\,0.0038]$  & $[-0.002,\,0.002]$  & $c_{24(8)}$ & $-$ & $-$ & $[-0.012,\,0.013]$
\\[2mm]
$c_{13(2)}$ & $[-0.12,\,0.076]$ & $[-0.11,\,0.073]$ & $[-0.06,\,0.038]$ & $c_{26(1)}$ & $-$ & $-$ & $[-0.006,\,0.007]$
\\[2mm]
$c_{13(4)}$ & $-$ & $[-2.2,\,3.4]$ & $[-0.038,\,0.06]$ & $c_{26(2)}$ & $-$ & $-$ & $[-0.006,\,0.007]$
\\[2mm]
$c_{14(1)}$ & $-$ & $[-0.58,\,1.1]$ & $[-0.02,\,0.04]$ & $c_{26(3)}$ & $-$ & $[-0.19,\,0.17]$ & $[-0.007,\,0.006]$
\\[2mm]
$c_{14(2)}$ & $-$ & $-$ & $[-0.04,\,0.02]$ & $c_{26(4)}$ & $-$ & $-$ & $[-0.007,\,0.006]$ 
\\[1mm]
\hline\hline
\end{tabular}
}
\caption{\sf Allowed ranges for both of the RH (upper row) and LRH operators (lower row), for $\bar{\epsilon}\sim 10^{-2}$ (3rd and 7th columns), $\bar{\epsilon}\sim 1$ (4th and 8th columns), compared with respect to those for the hierarchical case $\bar{\epsilon}\sim 10^{-4}$ suggested by the diboson excess (2nd and 6th columns). The ranges were obtained via the redefined coefficients in Table~\ref{Redefined-coefficients}, and implementing the 95\% CL ranges from EWPD bounds, and the 90\% CL limits from TGC, $hVV$-couplings and QGC constrains from Tables~\ref{S-T-bounds}-\ref{TGC-bounds}-\ref{hVV-bounds} and~\ref{QGC-bounds} respectively. One non-zero operator was assumed at a time.}
\label{All-RH-LRH-coefficients}
\end{table}

%

\providecommand{\href}[2]{#2}\begingroup\raggedright\endgroup

\end{document}